\documentclass[aps,prb,amsmath,showpacs,twocolumn,12p,floatfix]{revtex4-1}

\usepackage{graphicx,bm,enumerate,amssymb}
\usepackage{comment,color}

\newcommand{\ket}[1]{|#1\rangle}
\date{August 15, 2012}

\begin{document}

\title{Paired chiral spin liquid with a Fermi surface in $S=1$ model on the triangular lattice}

\author{Samuel Bieri}
\author{Maksym Serbyn}
\author{T.~Senthil}
\author{Patrick A.~Lee}

\affiliation{Department of Physics, Massachusetts Institute of Technology, 77 Massachusetts Avenue, Cambridge, Massachusetts 02139, USA}

\begin{abstract}
  Motivated by recent experiments on Ba$_3$NiSb$_2$O$_9$, 
  we investigate possible quantum spin liquid ground states for spin $S=1$ Heisenberg models on the triangular lattice. We use Variational Monte Carlo techniques to calculate the energies of microscopic spin liquid wave functions where spin is represented by three flavors of fermionic spinon operators. These energies are compared with the energies of various competing three-sublattice ordered states. Our approach shows that the antiferromagnetic Heisenberg model with biquadratic term and single-ion anisotropy does not have a low-temperature spin liquid phase. However, for an SU(3)-invariant model with sufficiently strong ring-exchange terms, we find a paired chiral quantum spin liquid with a Fermi surface of deconfined spinons that is stable against all types of ordering patterns we considered. We discuss the physics of this exotic spin liquid state in relation to the recent experiment and suggest new ways to test this scenario.
\end{abstract}


\maketitle

\section{Introduction}

Quantum spin liquids (QSL) are interesting states of matter with long-range entanglement that may exhibit exotic properties such as unbroken lattices symmetries at low temperature, quasiparticle fractionalization, emergent gauge fields, braid statistics, and chiral edge modes.\cite{lee08,balents10} The existence of QSL or resonating-valence-bond (RVB) states in two dimensions was first conjectured by Anderson as possible low-temperature phases of the spin-$1/2$ antiferromagnetic Heisenberg model on the triangular lattice.\cite{anderson72} Shortly after, a fascinating relation of RVB states with high-temperature superconductivity was uncovered: Upon doping, some quantum spin liquids are expected to give rise to unconventional superconductivity.\cite{anderson87,rmftReview04,leeReview08} So far, several experiments found indication of spin liquid behavior in a number of geometrically frustrated two-dimensional spin-$1/2$ antiferromagnets.\cite{kanoda03,yslee07,maegawa08,balicas11a} Spin systems with higher values of spin, however, usually show a strong tendency towards long-range ordering and lattice-symmetry breaking at low temperature. 

Last year, highly surprising experimental results\cite{balicas11b} found spin-liquid behavior in new structural phases of Ba$_3$NiSb$_2$O$_9$. In the so-called 6H-B phase, obtained through a high-pressure treatment of this antiferromagnetic insulator, the Ni$^{+2}$-ions, carrying effective spin $S=1$, arrange in presumably weakly coupled layers of triangular lattices. No magnetic ordering was observed down to 0.35~K despite a large Curie-Weiss temperature of $\Theta_{CW}\simeq -75$ K, and the magnetic susceptibility (after substraction of orphan spin contribution) was found to saturate at low temperature $T$. Furthermore, measurement of the magnetic specific heat found $C_M \propto T$. These properties are highly unusual for an insulator but are typical for metallic states. For example, spin-wave theory for conventional long-range ordered states predicts a specific heat $C_M \propto T^3$ at low temperature.\cite{ashkroftMermin}

So far, a number of theoretical attempts have been made to explain these experiments. Two possible spin-liquid candidates were proposed.\cite{serbyn11,xu12} 
In Ref.~[\onlinecite{xu12}], a representation of the spin $S=1$ operator in terms of {\it four} flavors of fermionic spinons and their possible mean-field states were conjectured. Such a fractionalization into four spinon flavors is most natural in the case of a two-orbital Hubbard model with not too strong interactions (Hund coupling) between the electrons. The minimal number of spinons required to represent spin $S=1$ is {\it three}.\cite{liuNg10a,liuNg10b} On the basis of this three-fermion representation, an exotic QSL state was proposed by some of us in Ref.~[\onlinecite{serbyn11}] that well reproduces the phenomenology of the experiment on Ba$_3$NiSb$_2$O$_9$. However, the energetic competitiveness of these spin liquid states in microscopic spin models was not investigated in those papers. Another scenario not involving spin-liquid states was recently proposed in Ref.~[\onlinecite{chen12}], where inter-layer couplings between the Ni$^{2+}$ spins tune the system to a quantum critical point. These authors predicted a $T$-linear specific heat in some temperature range.


Currently, the details of the effective spin model describing Ba$_3$NiSb$_2$O$_9$ are not known. In this paper, we will not propose a realistic microscopic spin model for this material. Instead, we want to investigate two families of promising antiferromagnetic triangular-lattice spin-one models at the variational level. The aim is to determine whether, variationally, the natural quantum spin-liquid candidates (involving three spinon flavors) have a chance to win over long-range ordered ground states in these microscopic models. First, we consider the bilinear-biquadratic Heisenberg model with single-ion anisotropy term. In this model, we do not find evidence for a low-temperature QSL phase. We further propose an SU(3) symmetric model with three-site ring-exchange terms. In this model, for strong ring-exchange terms, we find that an exotic spin liquid state is stabilized. We discuss the phenomenology of this state and propose further experimental tests of this scenario.
While the theory for $S=1/2$ QSL is well developed and has a long history, much less is known about spin liquids for $S=1$. Here we present new methods and results on this problem.\cite{noteS1}


This paper is organized as follows. In the next section, we introduce the representation of spin $S=1$ in terms of three fermionic spinon operators. Section~\ref{sec:wf} describes all spin liquid wave functions as well as the long-range ordered states that we are considering. In Sec.~\ref{sec:varres}, we introduce two spin models and present the variational results we found for these models.
In Sec.~\ref{sec:gaugetheory}, we discuss the low-energy field theories, and in Sec.~\ref{sec:edgemodes} the edge modes corresponding to the chiral d+id QSL state that we found to be stabilized in the ring-exchange model.
Section~\ref{sec:phenomenology} discusses the response function and other physical properties of this state and, finally, we conclude in Sec.~\ref{sec:conclude}.

\section{Spinon representation}\label{sec:spinrep}

To construct spin liquid states for spin $S=1$, we follow an approach similar to the one outlined in Ref.~[\onlinecite{liuNg10a}]. We write the spin operators in terms of three flavors of fermionic spinons, $f_a$, in the following way:
\begin{equation}\label{eq:spinrep}
  S_a = -i \varepsilon_{abc} f_b^\dagger f_c\,,
\end{equation}
where $a\in\{x,y,z\}$. In this paper, repeated indices are always summed over. We choose to work with operators $f_a$ that create spin states $|a\rangle$ in the time-reversal invariant basis, i.e.,
\begin{equation}\begin{split}\label{eq:xbasis}
 |x\rangle &= \frac{1}{\sqrt{2}}(|1\rangle + |\bar 1\rangle),\\
 |y\rangle &= \frac{i}{\sqrt{2}}(|1\rangle - |\bar 1\rangle),\\
 |z\rangle &= -i|0\rangle \,,
\end{split}\end{equation}
where $|1\rangle$, $|\bar 1\rangle$, and $|0\rangle$ are $S_z$-eigenstates with eigenvalues $\pm 1$ and $0$, respectively. 

By representing spin in terms of fermions, we have enlarged the Hilbert space. The fermion operators act in the eight-dimensional Fock space while the original spin space is three dimensional. In order to recover the physical subspace, a local constraint on the fermionic occupation number has to be enforced,
\begin{equation}\label{eq:constr}
n := \sum_a f_a^\dagger f_a \equiv N_f\, .
\end{equation}
Both particle ($N_f=1$) or hole ($N_f=2$) subspaces can be chosen. Furthermore, the spin operator remains invariant under the transformations
\begin{equation}\label{eq:u1}
  f_a \mapsto e^{i\phi} f_a
\end{equation}
and
\begin{equation}\label{eq:ph}
  f_a \mapsto f_a^\dagger\, .
\end{equation}
Equation~\eqref{eq:ph} is a {\it particle-hole} transformation and the constraint \eqref{eq:constr} is changed according to $N_f \mapsto 3 - N_f$. Hence, the local symmetry group for this representation of spin $S=1$ operators is the semi--direct product $U(1)\rtimes {\mathbb Z}_2$.\cite{liuNg10a}


In the time-reversal-invariant basis \eqref{eq:xbasis}, the quadrupolar operators, defined as $Q_{ab} = (S_a S_b + S_b S_a)/2 - 2/3\,\delta_{ab}$, acquire a particularly simple form.\cite{pencLauchli} In the particle representation ($N_f=1$), we have
\begin{equation}
  S_a S_b = \delta_{ab} - f_a^\dagger f_b\,,
\end{equation}
and $Q_{ab} = \delta_{ab}/3 - (f_a^\dagger f_b + f_b^\dagger f_a)/2$.

In order to analyze a particular spin $S=1$ lattice model within the spinon representation \eqref{eq:spinrep}, one may start by decoupling the spinon-interaction terms with the help of a Hubbard-Stratonovich transformation. To implement the constraint \eqref{eq:constr} and the symmetry properties \eqref{eq:u1} and \eqref{eq:ph} in this theory, a compact gauge potential for the local symmetry group has to be introduced in the path integral.\cite{leeReview08,leeLee05} This procedure enables derivation of a low-energy effective theory for possible spin liquid phases but does not address microscopic stability for a particular Hamiltonian. Better suited for this purpose is a variational wave function approach.
In this approach, unphysical states are removed by hand from wave functions that correspond to possible low-temperature phases of the theory. This allows the construction of a new class of genuine microscopic variational wave functions for spin-one models. Determining the best variational state for a spin model then provides guiding information about the low-temperature phase of
the model. In the present paper, we first follow this approach, and discuss possible microscopic Hamiltonians. We also use the effective field theory to discuss some properties of the proposed QSL states.

\section{Variational wave functions}\label{sec:wf}

In this section, we introduce two classes of microscopic variational wave functions for spin $S=1$ on the triangular lattice. First, we describe quantum spin liquid wave functions that do not break the space group symmetries of the lattice. Second, we outline a general approach for constructing competitive long-range ordered states that have an enlarged unit cell.

\subsection{Quantum spin liquid wave functions}\label{sec:wf:qsl}

We start by writing down quadratic ``trial'' Hamiltonians in terms of the spinon operators,
\begin{equation}\begin{split}\label{eq:Hqsl}
  H_\text{qsl} = \sum_{\langle i,j\rangle} \{ s f_{a i}^\dagger f_{a j} + \Delta^{a b}_{ij} f_{a i} f_{b j} + \text{h.c.}\}\\ - \mu_a \sum_j f_{a j}^\dagger f_{a j}\, .
\end{split}\end{equation}
The sum $\langle i,j\rangle$ goes over the nearest-neighbor links of the triangular lattice. In this trial Hamiltonian, the emergent gauge fields that would be present in the corresponding low-energy theory are omitted. Particular values for the mean-field parameters $s=\pm 1$, $\Delta^{ab}_{ij}$, and $\mu_a$ represent possible low-temperature QSL phases. $s=\pm 1$ corresponds to flux of $\pi$ or zero through all triangles of the lattice. Next, we assume unbroken spin-rotation symmetry around the $z$ axis, and we focus on $S_z^{\text{tot}}$ eigenstates with $S_z^{\text{tot}} = 0$. Note that under spin rotations, ${\bm f}_j = (f_{x}, f_{y}, f_{z})_j$ transform as real vectors [i.e.,\ ${\bm f}_j$ transform in the adjoint representation of SU(2)]. Furthermore, we restrict ourselves to states that have a single site per unit cell and that do not break the space group symmetries of the lattice (translations, rotations, and inversion). In this situation, the following cases exhaust the possible QSL candidates on the triangular lattice:
\begin{enumerate}[(i)]
  \item U(1) state: $\Delta^{a b}_{ij}=0$.
  \item Equal-flavor pairing: $\Delta^{zz}_{ij}\neq 0$, $\Delta^{xx}_{ij}=\Delta^{yy}_{ij} \neq 0$, $\Delta^{a b}_{ij}=0$ otherwise.
  \item $x\text{-}y$ pairing: $\Delta^{xy}_{ij} = -\Delta^{yx}_{ij} \neq 0$, $\Delta^{a b}_{ij} = 0$ otherwise. 
\end{enumerate}
The chemical potentials for $x$ and $y$ fermions are chosen to be identical, $\mu_x = \mu_y$. 
Other possible pairings $\Delta^{ab}_{ij}$ than the ones considered in (ii) or (iii) violate our symmetry requirements.\cite{noteSym}


On the one hand, for $\Delta^{xx}_{ij} = \Delta^{yy}_{ij} = \Delta^{zz}_{ij}$, equal-flavor pairing (ii) corresponds to spin-one {\it singlet} pairing. The pairing term in \eqref{eq:Hqsl} creates a state $({\bm f}_i^\dagger\cdot{\bm f}_j^\dagger)|\bar0\rangle$ that is invariant under spin rotation; hence, it is a singlet. In general, for $\Delta^{xx}_{ij} = \Delta^{yy}_{ij} \neq \Delta^{zz}_{ij}$, the state is not an eigenstate of $({\bm S}_{ij}^{\text{tot}})^2 = ({\bm S}_i + {\bm S}_j)^2$. However, for $\Delta^{xx}_{ij} = \Delta^{yy}_{ij} = -\Delta^{zz}_{ij}/2$ one can check that $({\bm S}_{ij}^\text{tot})^2 = 6$; therefore, this bond operator creates a spin-one {\it quintuplet}.
On the other hand, the $x\text{-}y$ pairing bond operator $\Delta^{ab}_{ij}f_{a i}^\dagger f_{b j}^\dagger$, (iii), creates a spin-one {\it triplet}. To see this, let us denote the state by $|1 \rangle_{ij} = (|x y\rangle_{ij} - |y x\rangle_{ij}) \propto (|1 \bar 1\rangle_{ij} - |\bar 1 1\rangle_{ij})$. Since $({\bm S}_{ij}^\text{tot})^2 = 4 + 2 {\bm S}_i\cdot{\bm S}_j$, and $[{\bm S}_i\cdot{\bm S}_j + 1 ] |1 \rangle_{ij} = 0$, we have $({\bm S}_{ij}^\text{tot})^2 = 2$.
Note that the total spin per site for all these QSL states is small in the thermodynamic limit. We have $\sqrt{\langle ({\bm S}^\text{tot})^2 \rangle}/N \sim 1/\sqrt{N}$ where $N$ is the number of sites and ${\bm S}^\text{tot} = \sum_j {\bm S}_j$.

Due to the anticommuting spinon operators, the pairing parameters $\Delta^{ab}_{ij}$ must have particular symmetry properties under inversion of the link direction $\langle i,j\rangle$: For equal-flavor pairing (ii), we have $\Delta_{ij}^{aa} = -\Delta_{ji}^{aa}$; i.e., the pairing is {\it odd} under space inversion. For $x\text{-}y$ pairing (iii), we have $\Delta_{ij}^{xy} = \Delta_{ji}^{xy}$; i.e., the pairing is {\it even} under space inversion. This is in contrast to $S = 1/2$ spin liquids, where singlet pairing is even while triplet pairing is odd under space inversion.

In order to obtain a microscopic variational QSL wave function, we take the ground state $|\psi_0\rangle$ of \eqref{eq:Hqsl} and apply the Gutzwiller projector $P_G(n_j=1)$, enforcing single occupancy on each site and thereby removing unphysical components. In this way we construct a genuine spin-one resonating-valence-bond (RVB) spin-liquid wave function, generalizing similar approaches to $S=1/2$ spin liquids.\cite{leeReview08} In this paper, we choose to work in the microcanonical formalism where the fermion number is held fixed; i.e., we project the wave function to a fixed total number of spinon flavors, $N_a = \sum_j n_{a j}$,
\begin{equation}\label{eq:projstate}
  |{\bm N}\rangle = P_{\bm N} P_G(n_j=1) |\psi_0\rangle\,,
\end{equation}
with ${\bm N} = (N_x, N_y, N_z)$. Since $N_x = N_y$ (to maintain spin-rotation symmetry around the $z$-axis) and from the local constraint we have $2 N_x + N_z = N$, where $N$ is the number of lattice sites ($N=12\times 12$ in most of our calculations).
Expectation values in RVB wave functions \eqref{eq:projstate} can be calculated numerically within Variational Monte Carlo (VMC) techniques.\cite{gros88}
More technical details on our numerical scheme are given in the appendices. 

The possible complex phases (pairing symmetries) of $\Delta^{ab}_{ij}$ in (ii) and (iii) are restricted by the rotation symmetries of the lattice: Let us denote the nearest-neighbor links of the triangular lattice by $\hat 1 = (1,0)$, and $\hat 2, \hat 3 = (\pm 1,\sqrt{3})/2$. For equal-flavor pairing (ii), the pairing symmetry can be real $f$-wave with $\Delta^{aa}_{\hat 1} = -\Delta^{aa}_{\hat 2} = \Delta^{aa}_{\hat 3}$, or complex $p_x + i p_y$-wave (p+ip), with $\Delta^{aa}_{\hat 1} = \Delta^{aa}_{\hat 2}\, e^{- i \pi/3} = \Delta^{aa}_{\hat 3}\, e^{- i 2 \pi/3}$. For $x\text{-}y$ pairing (iii), the possible pairing symmetries are extended $s$-wave with $\Delta^{xy}_{\hat 1} = \Delta^{xy}_{\hat 2} = \Delta^{xy}_{\hat 3}$ and $d_x + i d_y$-wave (d+id) with $\Delta^{xy}_{\hat 1} = \Delta^{xy}_{\hat 2}\, e^{-i 2\pi/3} = \Delta^{xy}_{\hat 3}\, e^{- i 4 \pi/3}$. Higher angular momenta would require spinon pairing between farther-neighbor sites, which we choose to exclude from the present study.\cite{noteXY}

Symmetry of the QSL states \eqref{eq:projstate} under lattice rotations forbids mixing of different types of pairing symmetries in the Hamiltonian \eqref{eq:Hqsl}. For example, lattice rotation symmetry is broken in a state where the $f_z$ spinon is paired with $f$-wave, and $f_x$, $f_y$ are paired with p+ip pairing symmetry. Similarly, in the $x\text{-}y$ paired QSL (iii), $f_z$ must remain unpaired unless lattice rotation symmetries are broken. The reason is the following: After performing a lattice rotation on the mean-field Hamiltonian \eqref{eq:Hqsl}, one would like to find a gauge transformation \eqref{eq:u1} that brings it back to the original form. If such a gauge transformation exists, then the corresponding spin wave function \eqref{eq:projstate} is unchanged by the rotation (after Gutzwiller projection and up to a phase). However, since all three spinon flavors transform with the same U(1) phase, such a gauge transformation can only exist when all spinon flavors have identical pairing symmetries.\cite{Wen02, noteMP}



The QSL states have the following properties: Extended $s$-wave and $f$-wave states respect parity $P$ (reflection on a symmetry axis of the lattice) and time-reversal symmetry $\Theta$. The p+ip and the d+id states, however, break both $P$ and $\Theta$, but conserve the product $\Theta P$. In this sense, they can be termed chiral spin liquids,\cite{wenWilcekZee89} albeit for spin $S=1$. The p+ip state is fully gapped and, therefore, a conventional topological state of matter. The d+id state, on the other hand, represents a new class of paired chiral states in two dimensions that exhibit both $P$- and $\Theta$-symmetry breaking {\it and} a gapless bulk Fermi surface at the same time. These exotic properties will be discussed in more detail below and in later sections.

In the U(1) spin liquid ($\Delta^{ab}=0$), all three spinon flavors have a Fermi surface. This corresponds to the Coulomb phase of the emergent U(1) gauge theory where the photons are massless. The paired states with $\Delta^{ab}\neq 0$ correspond to ``Higgs'' phases where the global U(1) symmetry is spontaneously broken and the photon acquires a mass.\cite{fradkin79} Among the equal-flavor pairing states, the $f$-wave state has gapless nodal points in the spectrum while the p+ip QSL is fully gapped. In the $x\text{-}y$ paired QSL, the spin excitations are gapped. However, the nematic ($S_z=0$) excitations form a gapless Fermi surface of weakly interacting (and therefore deconfined) spinons. We expect the Fermi surface to survive after Gutzwiller projection because the other fermion flavors are gapped and the U(1) gauge field is also gapped due to the Higgs mechanism. Specific heat and spin susceptibility of an $x\text{-}y$ paired (triplet) QSL are consistent with the recent experiments on Ba$_3$NiSb$_2$O$_9$.\cite{serbyn11}


The variational parameters we are using for the microscopic QSL wave functions are the amplitudes $|\Delta^{ab}|$ for all pairing symmetries discussed above and the chemical potentials $\mu_x$ and $\mu_z$. Furthermore, we consider the cases $s = \pm 1$ in \eqref{eq:Hqsl}, corresponding to the presence or the absence of $\pi$ flux through the triangles of the lattice. For the paired states, $N_z$ is used as an additional variational parameter (independent of $\mu_z$; see Appendix~\ref{app:fermionic} for more details). 

\subsection{Long-range ordered states}

In order to make reliable statements about the low-temperature phase of a spin model, the energies of QSL wave functions have to be compared with competitive long-range ordered states. Here, we consider natural ordering patterns that are suggested within a simple product-state ansatz (e.g., a 120$^\circ$ magnetic ordering in the case of the antiferromagnetic Heisenberg model on the triangular lattice). The QSL wave functions \eqref{eq:projstate} are highly correlated states. To be able to compare the variational energies, we also need to introduce nontrivial quantum correlations to the ordered states.

Here, we use the following two complementary schemes to introduce quantum corrections on top of long-range ordered product states. The first approach builds on the fermionic representation and gauge theory description of the spin model. 
Long-range ordered phases can be captured within the following quadratic trial Hamiltonian,
\begin{equation}\begin{split}\label{eq:Hord}
  H_\text{ord} =  s \sum_{\langle i,j\rangle} f_{a i}^\dagger f_{a j} - h \sum_j d^{a*}_{j} d^b_{j} f_{a j}^\dagger f_{b j}\\
  - \mu_a \sum_j f_{a j}^\dagger f_{a j}\, .
\end{split}\end{equation}
Similar to the QSL wave functions \eqref{eq:projstate}, the Gutzwiller-projected ground state of \eqref{eq:Hord} serves as a variational state. The normalized complex vectors ${\bm d}_j$ specify a particular spin-one ordering pattern. The variational parameter $h$ interpolates from the U(1) spin liquid ($h=0$) to the product state $|\psi_\text{p}\rangle = \prod_j \sum_a d^a_{j} |a\rangle_j$ when $h\rightarrow\infty$. As before, we set $\mu_x = \mu_y$; $\mu_x - \mu_z$ is taken as a variational parameter and we consider $\pi$- and 0-flux states by $s=\pm 1$.

Another route to constructing correlated long-range ordered wave functions is to apply spin Jastrow factors to a product state. The analysis of such wave functions for the spin-$1/2$ antiferromagnetic Heisenberg model on the triangular lattice was pioneered by Huse and Elser in Ref.~[\onlinecite{huseElser88}]. For that model, Huse-Elser wave functions were found to give low variational energies, comparable to exact energies on small clusters. A generalization of the Huse-Elser wave function to the case of spin $S=1$ can be written as
\begin{equation}\label{eq:jastrow}
  |\mathcal{J}\rangle = \exp( -\sum_{\langle i,j\rangle} \{ \beta (S_{z i} S_{z j}) + \gamma (S_{z i} S_{z j})^2 \}) |\psi_\text{p}\rangle\,.
\end{equation}
Here, $|\psi_\text{p}\rangle$ is a product state of spin one. 
In this paper, we restrict ourselves to nearest-neighbor Jastrow factors, and take $\beta$, $\gamma$ to be real variational parameters. 

A general spin-one product state can be written as
\begin{equation}\label{eq:prodstate}
  |\psi_\text{p}\rangle = \prod_j\sum_a d^a_{j} |a\rangle_j\,,
\end{equation}
where $|a\rangle \in \{|x\rangle,|y\rangle,|z\rangle\}$ span the local Hilbert space; see Eq.~\eqref{eq:xbasis}. Let us write ${\bm d} = {\bm u} + i {\bm v}$, where ${\bm u}$ and ${\bm v}$ are real vectors, and consider the single-site state $|\psi\rangle = \sum_a d^a|a\rangle$. We can always take ${\bm d} = (d^x,d^y,d^z)$ to be normalized and ${\bm u}\cdot{\bm v} = 0$ (choice of phase). The spin expectation value in this state is given by
\begin{equation}\label{eq:spinp}
  \langle {\bm S} \rangle = 2{\bm u}\wedge {\bm v}\, .
\end{equation}
If ${\bm d}$ is real, the corresponding state is a {\it spin nematic} with $\langle{\bm S}\rangle = 0$. In this case, ${\bm d}$ is called the {\it director} and we have
\begin{equation}\label{eq:dirp}
  \langle S_{a}^2\rangle = 1 - (d^{a})^2\, .
\end{equation}
On the other hand, ${\bm u}^2 = {\bm v}^2 = 1/2$ corresponds to a {\it spin coherent} state where $|\langle {\bm S} \rangle| = 1$ is maximal.\cite{pencLauchli}

The fermionic states \eqref{eq:Hord} and the Huse-Elser wave functions \eqref{eq:jastrow} are two quite general and complementary ways to introduce nontrivial quantum fluctuations on top of spin-one product states \eqref{eq:prodstate}.
Although additional variational parameters can be built into the product state itself, we need to choose a suitable (family of) product states (specified by ${\bm d}_j$) to start with. As the example by Huse and Elser\cite{huseElser88} suggested, good ground-states energies can be obtained by choosing the product states that minimize the energy of the spin model. Below we will use a similar choice.


\section{Models and variational results}\label{sec:varres}
\subsection{Bilinear-biquadratic Heisenberg model with single-ion anisotropy}

We start by considering the simplest extension of the spin-one Heisenberg antiferromagnet on the triangular lattice,
\begin{equation}\label{eq:KDmodel}
  H_{KD} = \sum_{\langle i,j\rangle} \{ {\bm S}_i \cdot {\bm S}_j + K ({\bm S}_i \cdot {\bm S}_j)^2 \} + D \sum_j S_{zj}^2\,,
\end{equation}
where we set the Heisenberg exchange energy $J=1$. In this study, we want to restrict ourselves to the parameter range $|K|\le 1.5$ and $|D|\le 1.5$.
For $D=0$, it is known that the ground state of this model exhibits $120^\circ$~antiferromagnetic order when $K<1$, and $90^\circ$~antiferro-nematic (also called antiferro-quadrupolar) order when $K>1$.\cite{lauchli06,tsunetsuguArikawa06} For large easy-axis anisotropy $D\gg 1$, the ground state is a ferro-nematic product state with $S_{z j} = 0$ on every site.

For intermediate values of $D$, one may expect an $x$-$y$ paired QSL to be stabilized in this model. Since $S_{zj}^2 = 1 - n_{zj}$, the single-ion anisotropy $D$ acts as a chemical potential for the $f_z$ spinon. For nonzero $D$, the Fermi surfaces of $f_z$ and of $f_x$, $f_y$ in the U(1) state are expected to mismatch, and it is conceivable that $f_x$ and $f_y$ pair while leaving $f_z$ with a spinon Fermi surface. Indeed, unconstrained mean-field theory in Ref.~[\onlinecite{serbyn11}] found that for $K\lesssim 0.5$, the p+ip QSL wins, while for $K\gtrsim 0.5$, the d+id state with a spinon Fermi surface is the most stable QSL candidate (However, in contrast to above intuition, the phase boundary showed only weak dependence on $D$). In the present paper we find that the variational energy of ordered states is always lower than the one of the QSL states when the local constraint \eqref{eq:constr} is taken into account exactly.


A variational study of the model~\eqref{eq:KDmodel} at the level of product states performed in Ref.~[\onlinecite{thesisToth11}] suggested a three-sublattice ordering pattern, generalizing the ordering pattern of Ref.~[\onlinecite{tsunetsuguArikawa06}] to $D\neq 0$. Motivated by this proposal, we choose the following two classes of three-sublattice product states as an input to our correlated ordered states discussed above. First, we consider an antiferromagnetic (AFM) state where the spins $\langle\psi_\text{p}| {\bm S}_j |\psi_\text{p}\rangle$ have a constant length and lie in a common plane at an angle of $120^\circ$ to each other on nearest-neighbor sites. The average spin length, $|\langle\psi_\text{p}| {\bm S}_j|\psi_\text{p}\rangle|$, is taken as a variational parameter. In the notation of Eqs.~\eqref{eq:Hord} and \eqref{eq:prodstate}, the spin states on sublattices $A$, $B$, and $C$ of the triangular lattice are written as
\begin{equation}\begin{split}\label{eq:magord}
{\bm d}_{j\in A} &= (0,-i \sin\eta, \cos\eta)\,,\\
{\bm d}_{j\in B,C} &= (\pm\frac{\sqrt{3} i }{2} \sin\eta, \frac{i}{2} \sin\eta, \cos\eta)\,,
\end{split}\end{equation}
where $\eta\in[0,\pi/2]$ is a variational parameter. Using Eq.~\eqref{eq:spinp}, one may check that this state corresponds to $120^\circ$~antiferromagnetic ordering in the $x\text{-}y$ plane with $|\langle\psi_\text{p}| {\bm S}_j|\psi_\text{p}\rangle| = \sin 2\eta$. We also consider the same ordering in the $x$-$z$ plane.\cite{noteJ} For $\eta=\pi/4$, each site is in a spin-coherent state; i.e., $|\langle {\bm S}_j\rangle| = 1$. The values $\eta\in\{0,\pi/2\}$ correspond to spin-nematic states with $\langle {\bm S}_j\rangle = 0$. For $\eta=0$, all directors point along the $z$ axis (ferro-nematic state), whereas for $\eta=\pi/2$, the directors on nearest-neighbor sites lie in a common plane at an angle of 120$^\circ$ ($120^\circ$~nematic state).

As a second ordering pattern, we consider spin-nematic (NEM) states with $\langle\psi_\text{p}| {\bm S}_j|\psi_\text{p}\rangle = 0$. The angle between the directors on different sublattices is constant and taken as a variational parameter (``umbrella'' configuration). More precisely, we take the following family of spin-nematic states,
\begin{equation}\begin{split}\label{eq:nemord}
{\bm d}_{j\in A} &= (0,-\sin\eta, \cos\eta),\\
{\bm d}_{j\in B,C} &= (\mp\frac{\sqrt{3}}{2} \sin\eta, \frac{1}{2} \sin\eta, \cos\eta)\,,
\end{split}\end{equation}
where the variational parameter $\eta$ controls the angle between the nematic directors on different sublattices. As before, the special value $\eta=0$ corresponds to a ferro-nematic, while $\eta=\pi/2$ is a $120^\circ$~nematic state. At the intermediate value $\sin\eta=\sqrt{2/3}$, the directors are perpendicular to each other on neighboring sites ($90^\circ$~antiferro-nematic state\cite{tsunetsuguArikawa06}).


\subsubsection*{Results}

\begin{figure}
\includegraphics[width=.5\textwidth]{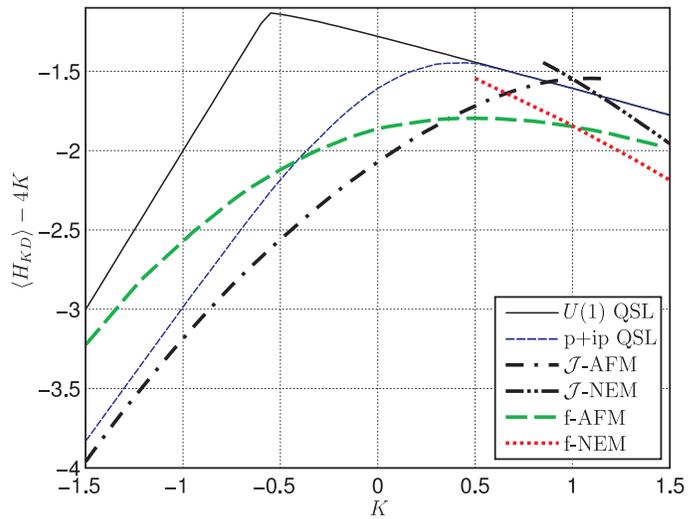}
\caption{Variational energies (per site) for the bilinear-biquadratic model, Eq.~\eqref{eq:KDmodel}, as a function of $K$, for $D=-0.4$. The system is $N=12\times 12$ lattice sites.}
\label{fig:KDmodel}
\end{figure}

Our variational results confirm the known phase diagram at $D=0$. Furthermore, we find that the three-sublattice ordering of the ground state persists for nonzero values of $D$; i.e., all QSL states are higher in energy than the three-sublattice ordered states we considered.\cite{noteD} A typical plot of the variational energies (for $D=-0.4$) is shown in Fig.~\ref{fig:KDmodel}. For $K\lesssim 0.3$, the magnetic Huse-Elser wave function ($\mathcal{J}$-AFM), Eq.~\eqref{eq:jastrow}, is the best variational state. For $0.3 \lesssim K \le 1$, the fermionic antiferromagnetic state (f-AFM), Eq.~\eqref{eq:Hord}, is the state with the lowest energy. As discussed above, the corresponding product states [specified in Eq.~\eqref{eq:magord}] are magnetically ordered with partially developed spins 
at $120^\circ$ angles between sublattices. For $D>0$, the ordered spins lie in the $x\text{-}y$ plane while for $D<0$, the spins order in a plane that contains the $z$ axis.
For $K\ge 1$, the fermionic nematic states (f-NEM) take over, with directors specified in Eq.~\eqref{eq:nemord}. For $D=0$, the best state is the $90^\circ$~antiferro-nematic state,\cite{tsunetsuguArikawa06} and for $D\neq 0$, the three nematic directors close ($D>0$) or open up ($D<0$) relative to the $z$ axis, depending on the sign of the single-ion anisotropy.  In the fermionic long-range ordered states \eqref{eq:Hord}, the optimal variational parameter is $h\simeq 1.5$. For this parameter value, the spinon excitations are fully gapped. Therefore, the spinons are confined\cite{noteConfinement} and we expect that bosonic spin-wave excitations for these ordered states capture the low-energy physics of this model.\cite{tsunetsuguArikawa06}

The energetically best quantum spin liquid states are the p+ip-state for $K\lesssim-1$ and $D\simeq 0$, and the unpaired U(1) state for $K\simeq 1$, both having zero flux through the triangles ($s=-1$). All the other QSL states show very small or no condensation energies with respect to the U(1) state. It is remarkable that for $K\simeq 1$, the U(1) state with three spinon Fermi surfaces is actually lower in energy than the optimized Huse-Elser wave function. We do not find any pairing instability of the U(1) spin liquid on the line $K\simeq 1$ for $D\gtrsim-0.8$. For $D \lesssim -0.8$, there is a small energy gain from pairing in the d+id channel. However, the ordered fermionic states are still lower in energy. When $D=0$, the three spinon Fermi surfaces match exactly. For $D>0$, the $f_z$ Fermi surface expands while $f_x$ and $f_y$ Fermi surfaces shrink. The opposite happens for negative $D$. The kink in the U(1) energy in Fig.~\ref{fig:KDmodel} marks the polarization to a ferro-nematic state with $\langle S_{z j}^2\rangle = 0$, for $K\lesssim -0.6$. That is, the spinon Fermi surfaces disappear at this point.

The variational energies for the spin-one Heisenberg antiferromagnet ($K = D = 0$) are displayed in Table~\ref{tab:heisenberg}. Note that the Heisenberg energy for the optimal product state of fully developed (coherent) spins ordered at 120$^\circ$ is $-1.5$. At the Heisenberg point, the spin liquids are even higher in energy than this uncorrelated product state.


At the point $K=1$ and $D=0$, the model \eqref{eq:KDmodel} enjoys an SU(3) symmetry\cite{pencLauchli} (see also next subsection). On the line $D=0$ and arbitrary $K$, the remaining symmetry is SO(3) spin-rotation. This symmetry is broken to U(1) (generated by $S_z$) when $D\neq 0$. However, on the line $K=1$ and arbitrary $D$, the model possesses an SU(2) symmetry generated by the operators $S_z$, $S_x^2 - S_y^2$, and $S_x S_y + S_y S_x$. The generator $S_x^2 - S_y^2$ allows rotation of the antiferromagnetic and the nematic ordered states [specified in Eqs.~\eqref{eq:magord} and \eqref{eq:nemord}] into each other and they are degenerate. This property of the product states remains valid after the introduction of quantum fluctuations via \eqref{eq:Hord} or \eqref{eq:jastrow}, and it explains the degenerate crossings for the ordered states seen in Fig.~\ref{fig:KDmodel} at $K=1$. See Appendix~\ref{app:sym} for a more detailed discussion of these symmetries.


\begin{table}
\begin{tabular}{l | c}
Variational state & Heisenberg energy\\
\hline
Huse-Elser $\mathcal{J}$-AFM & -1.783(1)\\
Fermionic f-AFM & -1.570(2)\\
p+ip spin liquid & -1.33(0) \\
U(1) spin liquid & -1.00(3)
\end{tabular}
\caption{Variational energies (per site) for the spin-one triangular-lattice Heisenberg antiferromagnet, \eqref{eq:KDmodel}, for $K = D = 0$; $N = 144$ sites.}\label{tab:heisenberg}
\end{table}

\subsection{SU(3) ring-exchange model}

In the last subsection we concluded that the simplest extension of the spin-one Heisenberg model, Eq.~\eqref{eq:KDmodel}, does not show quantum spin liquid behavior. To motivate another promising spin-one model, let us consider an SU(3) symmetric Hubbard model for three flavors of fermions $f_a$,
\begin{equation}\label{eq:su3hubbard}
  H_{SU(3)} = - t \sum_{\langle i, j\rangle} f_{a i}^\dagger f_{a j} + U \sum_j n_j^2\,, 
\end{equation}
where $n_j = \sum_a n_{a j} = \sum_a f_{a j}^\dagger f_{a j}$. Let us consider the case when each flavor is at $1/3$-filling ($\sum_j n_{a j}/N = 1/3$). For $U \gg |t|$, the low-energy subspace of this model corresponds to the spin-one Hilbert space. 
Similar to Refs.~[\onlinecite{granathOslund03},\onlinecite{macdonaldGirvinYoshioka88}], we can derive a low-energy effective spin-one Hamiltonian for \eqref{eq:su3hubbard}. To lowest order in $t$, we find the exchange term
\begin{equation}\label{eq:Uf0}
  \sum_{\langle i, j\rangle} f_{a i}^\dagger f_{b i} f_{b j}^\dagger f_{a j}\, .
\end{equation}
To next order, the following three-site term is expected to arise:
\begin{equation}\label{eq:Uf1}
  \sum_{\langle i, j, k\rangle} \{ f_{a i}^\dagger f_{b i} f_{b j}^\dagger f_{c j} f_{c k}^\dagger f_{a k} + \text{h.c.} \}\,,
\end{equation}
where the sum ${\langle i, j, k\rangle}$ is over elementary triangles of the lattice. Let us write the flavor exchange operators in \eqref{eq:Uf0} as $\mathcal{P}_{ij} = \sum_{a b} f_{a i}^\dagger f_{b i} f_{b j}^\dagger f_{a j}$. The three-site terms in \eqref{eq:Uf1} correspond to $\mathcal{P}_{ij}\mathcal{P}_{jk} + \mathcal{P}_{jk}\mathcal{P}_{ij}$. These operators move the local states clock- and anticlockwise around the triangles of the lattice.

In the case of a similar Hubbard model with {\it two} fermion flavors (spin $S=1/2$), the exchange operator $\mathcal{P}_{ij}$ appearing in the low-energy model corresponds to the Heisenberg term in spin language, $\mathcal{P}_{ij} = 2{\bm S}_i\cdot {\bm S}_j+1/2$. In this case, a three-site term $\mathcal{P}_{ij}\mathcal{P}_{jk} + \text{h.c.}$ is trivial in the sense that it can be reduced to a sum of two-site terms. For three flavors, however, the situation is different. In that case and spin $S=1$, one finds\cite{zhangWang06}
\begin{equation}\label{eq:exchange}
  \mathcal{P}_{ij} = {\bm S}_i \cdot {\bm S}_j + ({\bm S}_i \cdot {\bm S}_j)^2 - 1\,.
\end{equation}
Therefore, the lowest-order term \eqref{eq:Uf0} corresponds to the $KD$-model \eqref{eq:KDmodel} with $K=1$ and $D=0$. The next-order ring-exchange term \eqref{eq:Uf1} is a nontrivial perturbation since it cannot be reduced to two-site terms. Ring-exchange models for spin $1/2$
with nontrivial four-site plaquette terms\cite{thouless65} are believed to exhibit spin-liquid ground states.\cite{lauchli05,misguich99,montrunichTr}

Motivated by the three-flavor Hubbard model, we propose to study the SU(3) symmetric ring-exchange model,
\begin{equation}\label{eq:ringModel}
  H_{\alpha} = \cos\alpha \sum_{\langle i, j\rangle} \mathcal{P}_{ij} + \sin\alpha \sum_{\langle i,j,k\rangle}\{ \mathcal{P}_{ij} \mathcal{P}_{jk} + \text{h.c.} \}\, .
\end{equation}
The sum in \eqref{eq:ringModel} goes over nearest-neighbor links $\langle i, j\rangle$ and elementary triangles $\langle i,j,k\rangle$ of the lattice. The parameter of this model is $\alpha\in [-\pi,\pi]$.

\subsubsection*{Results}

\begin{figure}
\includegraphics[width=.5\textwidth]{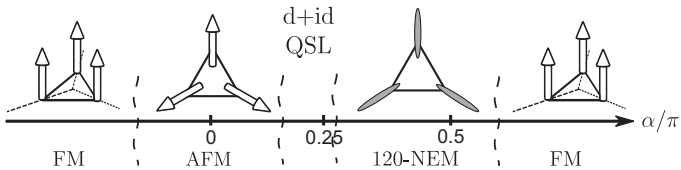}
\caption{Pictorial presentation of the variational phase diagram that we find for the SU(3) ring-exchange model~\eqref{eq:ringModel}.}
\label{fig:ringpic}
\end{figure}

An analysis of the model \eqref{eq:ringModel} in terms of general three-sublattice product states reveals a ferromagnetic phase in the parameter range $\alpha < -\arctan(3/4) \simeq-0.2\pi$ and $\alpha>\pi-\arctan(1/2) \simeq 0.85\pi$. Any uniform product state $\prod_j |a\rangle_j$ is an exact eigenstate with energy $\epsilon(\alpha) = 3\cos\alpha+4\sin\alpha$, and it is the lowest-energy three-sublattice product state in this parameter range. For $-\arctan(3/4) < \alpha < \arctan(3/2) \simeq 0.31\pi$ the $120^\circ$~antiferromagnetic product state is stabilized. Finally, in the range $\arctan(3/2) < \alpha < \pi - \arctan(1/2)$, the three nematic directors order in a common plane at an angle of $120^\circ$ to each other on nearest-neighbor sites (see Fig.~\ref{fig:ringpic}).

The above analysis with uncorrelated three-sublattice product states revealed the same ordering patterns as we found in the case of the $KD$-model investigated in the previous subsection. Therefore, we can use the same trial wave functions specified in Eqs.~\eqref{eq:magord} and \eqref{eq:nemord} to construct correlated ordered states for the ring-exchange model.


\begin{figure}
\includegraphics[width=.5\textwidth]{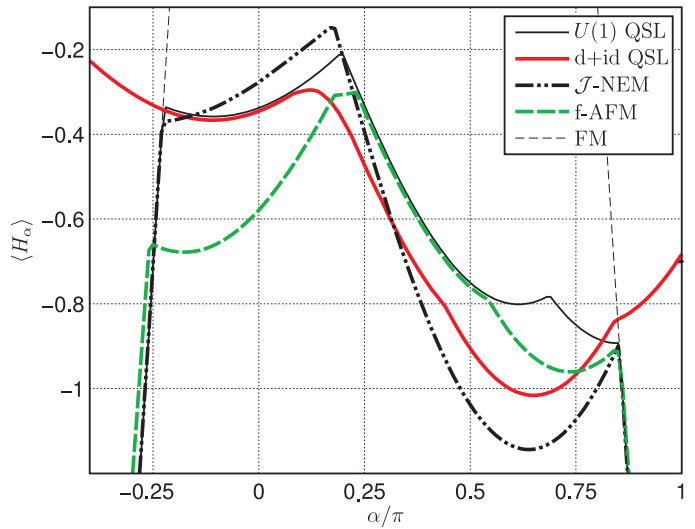}
\caption{Variational energies (per site) of the SU(3) ring-exchange model, Eq.~\eqref{eq:ringModel}, as a function of $\alpha/\pi$. $N=12\times 12$ lattice sites.}
\label{fig:Rex}
\end{figure}

\begin{figure}
\includegraphics[width=.5\textwidth]{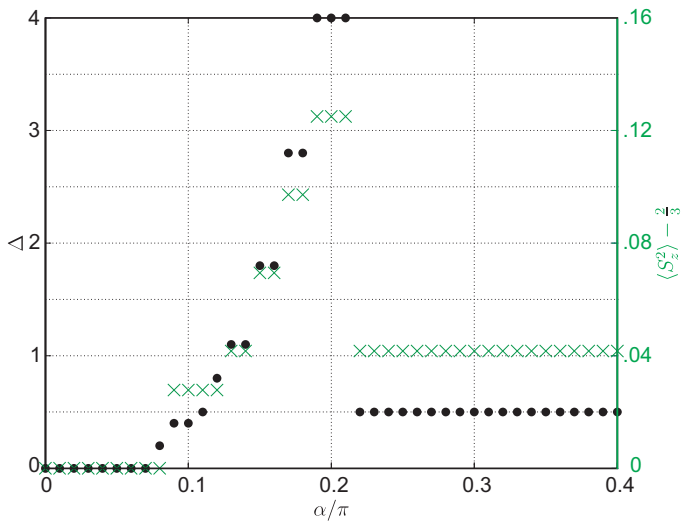}
\caption{Optimized variational parameters $\Delta = |\Delta^{xy}|$ (dot symbols, left scale) and $\langle S_z^2\rangle - 2/3$ ($x$ symbols, right scale) for the d+id QSL state in the ring-exchange model \eqref{eq:ringModel}. Among the states we consider, the d+id state has the lowest energy in the range $0.17\pi\lesssim\alpha\lesssim 0.33\pi$. For $0.17\pi\lesssim\alpha\lesssim 0.22\pi$, the optimal state is a $0$-flux state with $s=-1$; for $0.22\pi\lesssim\alpha\lesssim 0.33\pi$, we find a $\pi$-flux state with $\Delta\simeq 0.5$ and $s=1$.}
\label{fig:didpar}
\end{figure}

We calculate the variational energies of the QSL states \eqref{eq:Hqsl} as well as the energies of correlated three-sublattice ordered states \eqref{eq:Hord} and \eqref{eq:jastrow} for the ring-exchange model specified in Eq.~\eqref{eq:ringModel}. The results are presented in Fig.~\ref{fig:Rex}. We see that the conclusions we draw from the simple product state calculation above agree with the result using correlated wave functions in most of the parameter range. However, in the region between the AFM and the 120$^\circ$-nematic phase, around $\alpha\simeq\pi/4$, we find an extended region where the d+id QSL has the lowest energy (see also Fig.~\ref{fig:ringpic} for a scheme of the phase diagram). The optimal d+id variational parameter $|\Delta^{xy}|$ along with $\langle S_z^2\rangle - 2/3 = 1/3 - N_z/N$ are shown in Fig.~\ref{fig:didpar}. The ring-exchange term favors a $\pi$-flux d+id state with $s=1$: As $\alpha$ increases, the $0$-flux state with a large pairing term ($|\Delta^{xy}| \simeq 4$) changes to a $\pi$-flux state with $|\Delta^{xy}| \simeq 0.5$ at $\alpha\simeq 0.22\pi$.

Note that the d+id QSL phase in Fig.~\ref{fig:ringpic}, as well as the adjacent $120^\circ$~nematic phase, exhibit a ferro-quadrupolar order. For the d+id state this is apparent from Fig.~\ref{fig:didpar} since $\langle S_z^2\rangle>2/3$. In contrast, lattice rotation symmetry is unbroken in the d+id QSL while both adjacent ordered phases spontaneously break lattice rotation.\cite{refSU2}

As discussed above, for $\alpha=0$ and up to a constant, \eqref{eq:ringModel} corresponds to the model \eqref{eq:KDmodel} with $K=1$ and $D=0$. The ground state of this model was recently approached with density matrix renormalization group (DMRG) calculations in Ref.~[\onlinecite{bauer12}]. In this work, the authors found a three-sublattice ordering pattern that is consistent with our result. The DMRG energy is displayed in Table~\ref{tab:su3model} along with the variational energies of the lowest-energy states used in the present paper.

\begin{table}
\begin{tabular}{l | c}
State & SU(3) energy\\
\hline
Fermionic f-AFM & -0.57(8)\\
Huse-Elser $\mathcal{J}$-AFM & -0.27(7)\\
U(1) spin liquid & -0.34(3) \\
DMRG\cite{bauer12} ($N=8\times 10$)& -0.678
\end{tabular}
\caption{Variational energies for the SU(3) model, Eq.~\eqref{eq:ringModel}, at $\alpha=0$ on $N=12\times 12$ sites.}\label{tab:su3model}
\end{table}

We also consider additional perturbations to the ring-exchange model~\eqref{eq:ringModel} in order to assess the effect of such terms on possible low-temperature QSL phases. First, we add a single-ion anisotropy term $D \sum_j S_{zj}^2$. As discussed in the previous section, such a term breaks the SU(3) symmetry of the model to SU(2). For small $D$, the ordering plane is explicitly chosen. Large $D$ deforms the three-sublattice ordering pattern in a way similar to the bilinear-biquadratic Heisenberg model~\eqref{eq:KDmodel}. However, we find that the phase boundaries of the d+id state with the adjacent ordered states are barely affected by $D$ (we investigate the range $|D|\lesssim 1.5$). 

Second, we add a next-neighbor exchange term $J_2\sum_{\langle\langle i,j\rangle\rangle} \mathcal{P}_{ij}$ to \eqref{eq:ringModel}. Such a term strongly frustrates three-sublattice ordering. At $\alpha=0$, we find that the three-sublattice ordering is destroyed for $J_2$ as small as $J_2\simeq 0.25$, and the U(1) QSL has the lowest energy among our ansatz wave functions.
However, an analysis of this model in terms of product states reveals that the competing ordering pattern is {\it spiral}. So far, we have not included spiral states into our variational analysis and we reserve a detailed study for future work.

In conclusion, we find a region in the phase diagram of the SU(3) ring-exchange model \eqref{eq:ringModel} where a chiral d+id quantum spin liquid state is stabilized. The question remains whether this spin model can describe the relevant magnetic interactions in the 6H-B structure of Ba$_3$NiSb$_2$O$_9$. A perturbative expansion in $t/U$ of a two-band Hubbard model with an additional orbital degree of freedom and strong Hund coupling would produce a spin $S=1$ Heisenberg term ${\bm S}_i\cdot{\bm S}_j$ to order $t^2$. Only to next order, $t^4$, one expects biquadratic terms $({\bm S}_i\cdot{\bm S}_j)^2$ as well as three-site terms $({\bm S}_i\cdot{\bm S}_j)({\bm S}_j\cdot{\bm S}_k)$.\cite{bastardis07,fazekasBook,kittel60} Those three-site terms are present in our ring-exchange model, but we need them to be of the same order as the Heisenberg term. We found that a dominant nearest-neighbor Heisenberg term is detrimental to the stability of the d+id quantum spin liquid. Further terms in \eqref{eq:ringModel} like $({\bm S}_i\cdot{\bm S}_j)^2({\bm S}_j\cdot{\bm S}_k)$ would only come to order $t^6$ or higher in a perturbative expansion. At present, we cannot conclude if these higher order terms are relevant for the stability of the d+id QSL or not. 

While it is unclear, at present, whether the ring-exchange model \eqref{eq:ringModel} can realistically describe the spin-liquid phase of Ba$_3$NiSb$_2$O$_9$, it is a very natural model to study if one starts from an integer-filled three-band Hubbard model.
Such three- (and higher-) band Hubbard models are currently of great interest, both theoretically and experimentally, in the cold-atom community; see, e.g., Refs.~[\onlinecite{bauer12},\onlinecite{hermele10}].



\section{Gauge theory for the $\text{d+id}$ QSL}\label{sec:gaugetheory}


In this section, we propose to discuss the low-energy gauge theory description of the d+id spin liquid and some of its properties. In order to impose the local particle-number constraint \eqref{eq:constr}, the Lagrange multiplier $\lambda_{j}$ is introduced in the Euclidean path integral for the spinon partition function,\cite{leeReview08}
\begin{equation}\begin{split}\label{eq:partition}
  Z = &\int D\lambda \prod_a [Df_a^* Df_a]\, e^{ - S }\,.
\end{split}\end{equation}
The action is given by
\begin{equation}\label{eq:action}
  S = \sum_j \int_0^\beta d\tau\, \{f_{a j}^* (\partial_\tau - i \lambda_{j}) f_{a j} + i \lambda_{j} z_j  + H\}\, . 
\end{equation}
The Ising variables $z_j\in\{1,2\}$ specify whether site $j$ is constraint to the particle ($N_f=1$) or hole ($N_f=2$) representation. The Lagrange multiplier $\lambda_{j}$ turns out to be the temporal component of the emergent U(1) gauge field. $H$ is the microscopic spin-one Hamiltonian under consideration, written in terms of the spinon variables $f_{a j}$. As already stated in Eqs.~\eqref{eq:u1} and \eqref{eq:ph}, a local transformation that leaves the spin operator \eqref{eq:spinrep} invariant is given by an element $g_j\in U(1)\rtimes \mathbb{Z}_2$. Under the transformation $g_j=(e^{i\phi_j},\chi_j=\pm)$, the fields transform as
\begin{equation}\begin{split}\label{eq:ftransform}
  f_{a j} &\mapsto e^{i\phi_j} \left( \frac{1+\chi_j}{2} f_{a j} + \frac{1-\chi_j}{2} f^*_{a j} \right)\,,\\
  z_j &\mapsto 3 \frac{1 - \chi_j}{2} + \chi_j z_j\,,\\
\end{split}\end{equation}
and
\begin{equation}\begin{split}\label{eq:lambdatransf}
  \lambda_{j} &\mapsto \chi_j\lambda_{j} + \partial_\tau \phi_j\,.
\end{split}\end{equation}
Note that the action \eqref{eq:action} is not invariant under time-dependent particle-hole transformations $\chi_j$. Therefore, the particle-hole part of the local symmetry group is not a genuine gauge symmetry of the action. In the following, we can simply choose a particular {\it static} $\mathbb{Z}_2$ configuration, e.g.\ $z_j=1$, in \eqref{eq:action}. Furthermore, generic mean-field decouplings (see below) break this local particle-hole symmetry.



In a next step, the spinon interaction terms in $H$ can be decoupled by appropriate Hubbard-Stratonovich fields as is done in the usual slave particle formalism.\cite{leeReview08} To maintain the gauge invariance of the action, we need to introduce link variables $a_{ij}$ that are the space components of the U(1) lattice gauge field. The gauge field $(\lambda_j, a_{ij})$ mediates the interaction between the fermionic spinons. So far, all manipulations are formal transformations that do not change the physical content of the action. The question remains whether the resulting U(1) lattice gauge theory exhibits a phase with deconfined spinons. Possible low-temperature phases of the gauge theory are specified at the mean-field level by quadratic Hamiltonians \eqref{eq:Hqsl} and \eqref{eq:Hord}. 

Let us now specialize to the d+id phase that we found previously in the SU(3) ring-exchange model~\eqref{eq:ringModel}. This is a Higgs phase where particle-number conservation is spontaneously broken and the U(1) gauge field acquires a mass $m_0 \propto|\Delta^{xy}|$. At the same time, the fermions $f_x$ and $f_y$ are gapped and can safely be integrated out in the path integral. This generally leads to a Maxwell term for the U(1) gauge field in the low-energy effective action. Another low-energy term is a Chern-Simons term $\epsilon^{\mu\nu\lambda} a_\mu\partial_\nu a_\lambda$. The Chern-Simons term violates time reversal $\Theta$ and parity $P$. Therefore, its coefficient $\sigma_h$ cannot vanish in the chiral d+id spin liquid.\cite{wenWilcekZee89} Hence, in the continuum limit, we arrive at the following effective action
\begin{equation}\begin{split}
  S_{\text{eff}} = &\int d\tau d^2x\, \{ f_z^* [\partial_\tau - i a_0 - \mu_z + \frac{({\bm \nabla} - i{\bm a})^2}{2 m}] f_z\\
   &+ \frac{m_0}{2} a_\mu^2 + \frac{\sigma_h}{2} \epsilon^{\mu\nu\lambda} a_\mu\partial_\nu a_\lambda + \ldots \}\,,
\end{split}\end{equation}
where the ellipsis denotes higher-order terms in derivatives and gauge fields. 
In this theory, the $f_z$ spinon maintains its Fermi surface, and it is only weakly interacting via the massive photon. The excitations corresponding to $f_z$ are therefore deconfined in this phase.



\section{Chiral edge modes for the $\text{d+id}$ QSL}\label{sec:edgemodes}


In Ref.~[\onlinecite{volovik97}], it was shown that the d+id superconductor is a topological state with Chern number equal to two. From the bulk-edge correspondence, this indicates the presence of two chiral edge modes. A semiclassical argument\cite{senthilMarstonFisher99} supports this conclusion. In the next subsection, we recapitulate this semiclassical argument and generalize it to chiral topological superconductors. In subsection~\ref{sec:edgemodesB}, we specialize to the d+id QSL state. We calculate its energy spectrum on a triangular-lattice strip, and we discuss the corresponding low-energy edge theory.

\subsection{Edge modes in chiral superconductors}

In the bulk of a superconductor involving {\it two} fermion flavors, writing $\psi = (c_1, c_2^\dagger$), the Bogolubov equations are
\begin{equation}
\left(
  \begin{array}{cc}
    \xi_{\bm k}-E_{\bm k} & \Delta({\bm k})\\ 
    \Delta({\bm k})^* & -\xi_{\bm k}-E_{\bm k}\\ 
  \end{array}
\right) \psi_{\bm k} = 0\, .
\end{equation}
Here, we consider fully gapped superconductors with $|\Delta({\bm k})|>0$. The spectrum is given by $E_{\bm k} = \pm \sqrt{\xi_{\bm k}^2 + |\Delta({\bm k})|^2}$. 

Next, consider a superconductor with a boundary along the $x$-direction. Asymptotically (i.e., for $|{\bm k}|y \gg 1$), an incident bulk wave packet $\psi_{\bm k} e^{i {\bm k}\cdot {\bm r}}$ with ${\bm k} = (k_x, k_y)$ is reflected at the boundary to an outgoing wave packet $\psi_{{\bm k}'} e^{i {\bm k}'\cdot {\bm r}}$ with wave vector ${\bm k}' = (k_{x}, - k_{y})$.
The two wave packets ``see" the gap functions $\Delta({\bm k})$ and $\Delta({\bm k}')$, respectively. For a given incident wave vector ${\bm k}$, it seems therefore possible to map this problem on the half plane to a one-dimensional scattering problem where the order parameter $\Delta(y)$ interpolates from $\Delta({\bm k})$ as $y \rightarrow -\infty$ to $\Delta({\bm k}')$ as $y\rightarrow+\infty$:
\begin{equation}\label{eq:1dscatt}
\left(
  \begin{array}{cc}
     - i\partial_y - E & \Delta(y)\\ 
    \Delta(y)^* & i \partial_y - E\\ 
  \end{array}
\right) \psi_{k_x}(y) = 0\, .
\end{equation}
This one-dimensional problem can now be solved in the usual way.\cite{senguptaDasSarma01, stoneRoy04} For simplicity, let us choose a bulk potential of the form $\Delta({\bm k}) = \Delta e^{i l \theta({\bm k})}$ where $l\in\mathbb{Z}$ is the winding of the phase of the order parameter around the Fermi surface, and $\cos\theta({\bm k}) = k_x/|{\bm k}|$. A scattering state with incident angle $\theta$ results in an outgoing angle $\theta' = \pi - \theta$. The asymptotic potentials can therefore be chosen as $\Delta({\bm k}) = \Delta e^{i l (\theta-\pi/2)}$ and $\Delta({\bm k}') = \Delta e^{-i l (\theta - \pi/2)}$. Accordingly, the order parameter $\Delta(y)$ in \eqref{eq:1dscatt} has a constant real part $\Delta\cos l(\theta - \pi/2)$, and an imaginary part $\Delta\sin l(\theta - \pi/2)$ that changes sign across the boundary. This problem can be solved exactly for certain special cases of the interpolating gap profiles.\cite{takayamaLinMaki79, senguptaDasSarma01} For example, a kink profile with $\Delta(y) = \Delta [ \cos l(\theta - \pi/2) - i \tanh(y) \sin l(\theta - \pi/2) ]$ yields the bound-state spectrum
\begin{equation}\label{eq:boundstateenergy}
  E_{\theta} = \Delta \cos [ l (\theta - \frac{\pi}{2})]\, \text{sgn} [\sin l (\theta - \frac{\pi}{2}) ]\,,
\end{equation}
with corresponding eigenvectors
\begin{equation}\label{eq:boundstates}
  \psi_{\theta}(y) = \frac{1}{2 \cosh(y)}\, (1, \text{sgn} [\sin l (\theta - \frac{\pi}{2}) ])\, .
\end{equation}
We observe that $E_{\theta}$ as a function of $\theta$ vanishes exactly $|l|$ times. Therefore, there are $|l|$ gapless edge modes.
Using $\cos\theta \simeq k_x/k_f$ and expanding Eq.~\eqref{eq:boundstateenergy} around a node at momentum $k_x^n$, we get $E_{\theta} \simeq -l (k_x-k_x^n) \Delta/ k_f~+~\ldots$. The edge modes are chiral and propagate with the velocity $v_n = -l\Delta/ k_f$. In the simplest nontrivial and well-known case of a p+ip-superconductor\cite{stoneRoy04,senguptaDasSarma01,ivanov04} ($l=1$), a single chiral edge mode is located at $k^n_x = 0$. Higher angular momenta have chiral modes at $k^n_x \neq 0$. Since the phase winding of the order parameter around the Fermi surface is a topological property, we expect that the number of chiral edge modes is a robust feature of the state, too. The precise location of the nodes $\{k^n_x\}$ and the corresponding propagation speeds $|v_n|$, however, depend on further microscopic details.

\subsection{Low energy edge theory for the d+id QSL state}\label{sec:edgemodesB}

\begin{figure}
\includegraphics[width=.5\textwidth]{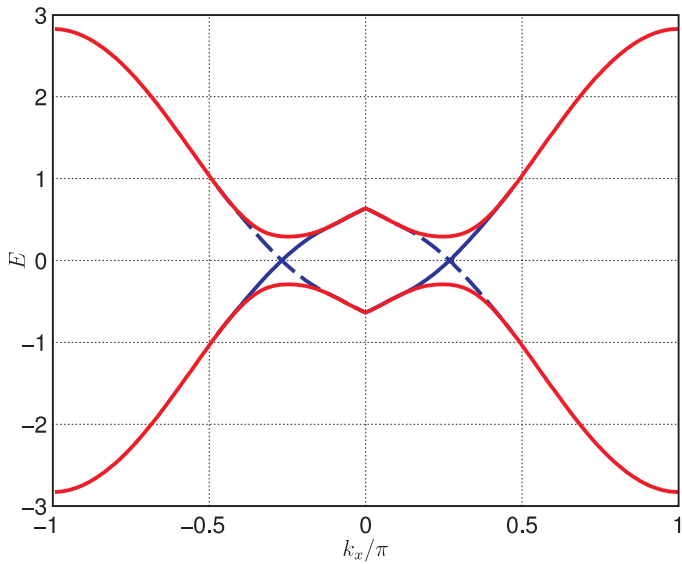}
\caption{The four lowest energy levels of the d+id mean-field state \eqref{eq:Hqsl} on an infinite triangular-lattice strip as a function of wave vector $k_x$ along the strip. The width of the strip is 200 sites. The boundaries are chosen to be parallel to one lattice direction and we use open boundary conditions. The spectrum of the $f_z$ spinon with a bulk Fermi surface is omitted. 
The gapless states (blue online) are localized on the lower boundary for left movers (dashed line), and on the upper boundary for right movers. The higher states (red online) are delocalized and the energy levels above them are ``dense".}
\label{fig:did_stripe}
\end{figure}

According to the above semiclassical argument, the d+id QSL state ($l=2$) is expected to exhibit two chiral edge modes located at wave vectors $k^n_x \simeq \pm k_f/\sqrt{2}$. To substantiate this claim, we calculate the spectrum of the d+id state \eqref{eq:Hqsl} on an triangular-lattice strip of infinite length [here, we neglect the local constraint \eqref{eq:constr} and work in the fermionic Fock space]. The four lowest energy levels are shown in Fig.~\ref{fig:did_stripe} as a function of wave vector $k_x$ along the strip. The triangular-lattice d+id state indeed exhibits two gapless left movers localized on the lower boundary and two right movers localized on the upper boundary. The spectrum of $f_z$ spinons with a bulk Fermi surface is omitted in Fig.~\ref{fig:did_stripe}.

As discussed above, the low-energy degrees of freedom localized on the edge for the d+id QSL state are two chiral Dirac fermions. To discuss the physics of these edge states, it is convenient to go to the spinon basis creating $S_z$ eigenstates. We have 
\begin{equation}\begin{split}
f_\sigma = \frac{1}{\sqrt{2}} (f_x - i \sigma f_y)\,,
\end{split}\end{equation}
with $\sigma=\pm 1$. We also denote $f_{\bar 1} = f_{-1}$. The $x\text{-}y$ (triplet) pairing term of the d+id state is $f_{x i} f_{y j} - f_{y i} f_{x j} = i(f_{1 i} f_{\bar 1 j} - f_{\bar 1 i} f_{1 j})$. Let us consider an edge along the $x$ direction and denote the momentum along the edge by $k = k_x\in [-\pi,\pi]$. The two gapless points in the boundary spectrum are denoted by $k_x^n = \pm k^0$ with $k^0>0$.

Using the semiclassical expression \eqref{eq:boundstates}, the edge states are created by operators
\begin{equation}\begin{split}\label{eq:modes}
  \chi_{\sigma}(k) & \sim f_{\sigma k} + \sigma \text{sgn}(k) f_{\bar\sigma -k}^\dagger\,,
\end{split}
\end{equation}
for $|k|\simeq k^0$. The excitations $\chi_{1}(k)$ and $\chi_{\bar 1}(k)$ carry spin $S_z=\pm 1$, respectively. Note that the edge states at positive and negative momenta $k$ are not independent: We have $\chi_{\sigma}^\dagger(k) = \sigma\text{sgn}(k)\,\chi_{\bar\sigma}(-k)$. The low-energy effective edge Hamiltonian is therefore given by
\begin{equation}
  \mathcal{H} = v_0 \sum_{k\simeq k^0, \sigma}  (k - k^0)\, \chi_{\sigma}^\dagger(k)\chi_{\sigma}(k)\,, 
\end{equation}
where the sum over $k$ is restricted to the vicinity of the node at momentum $+k_0$ to avoid double counting of states.



Similar to the ordinary quantum Hall effect, the chiral edge modes are expected to be robust with respect to disorder and impurities because no backscattering is possible.\cite{wen91} Furthermore, due to $S_z$ conservation, hybridization terms such as $f_z^\dagger \chi_\sigma$ cannot appear in the low-energy Hamiltonian. In a mean-field decoupling, interaction terms such as $f_z^\dagger f_z \chi_\sigma^\dagger\chi_\sigma$ only shift the chemical potentials of bulk and edge gapless modes, and do not significantly alter the edge physics. The presence of protected chiral edge modes carrying spin $S_z=\pm 1$ implies a quantized spin Hall conductivity. We also expect a thermal Hall conductivity in the d+id QSL state.

In the d+id QSL phase with unbroken lattice symmetries, $f_z$ must necessarily form a spinon Fermi surface (see Sec.~\ref{sec:wf:qsl}). However, this argument becomes invalid when the lattice symmetries are explicitly broken. For example, close to the boundary of the sample, symmetry allows a pairing term for $f_z$. Similarly, we expect the spinon to acquire a local gap in the vicinity of bulk impurities. This property makes the spinon Fermi surface hard to detect in any experiment that involves local probes.

\section{Response functions and physical properties of the $\text{d+id}$ QSL}\label{sec:phenomenology}

In this section, we release the local constraint \eqref{eq:constr} in order to analyze the spectral properties of the d+id mean field state. This can be justified from the point of view of the U(1) gauge theory since we are in a Higgs phase where gauge fluctuations can be neglected. In this case, the $f_z$ spinon can be treated as a weakly interacting Fermi liquid.
\subsection{Static spin susceptibility and NMR relaxation rate}

The response function, $R_{aa}(i \omega) = \sum_{ij}\int_0^\beta d\tau\, e^{ i\omega \tau} \langle S_{ai}(\tau) S_{aj}\rangle$, in the d+id state has the following properties: Since $f_x$ and $f_y$ fermions are paired, we have $R_{zz}(i\omega)=0$. However, $R_{xx}(i\omega) = R_{yy}(i\omega)$ do not vanish at low temperature. In the low-frequency, low-temperature limit, $|\omega| \ll T \rightarrow 0$, we find
\begin{equation}\label{eq:chilowT}
  \chi_{x}^0 = \text{Re}[ R_{xx}(0) ] = \int_{\text{BZ}} \frac{d^2k}{2\pi}\, \frac{E_{\bm k} - \text{sgn}(\xi^z_{\bm k}) \xi_{\bm k}^x}{E_{\bm k} (E_{\bm k} + |\xi^z_{\bm k}|)}
\end{equation}
where $\xi_{\bm k}^a = 2 s [ \cos({\hat 1}\cdot{\bm k}) + \cos({\hat 2}\cdot{\bm k}) + \cos({\hat 3}\cdot{\bm k})] - \mu_a$ is the dispersion of the $x$- and $z$-fermions. $E_{\bm k} = \sqrt{(\xi_{\bm k}^x)^2 + |\Delta^{xy}_{\bm k}|^2}$, and the d+id gap function is $\Delta^{xy}_{\bm k} = \Delta [ \cos({\hat 1}\cdot{\bm k}) + e^{2\pi i/3} \cos({\hat 2}\cdot{\bm k}) + e^{-2\pi i/3} \cos({\hat 3}\cdot{\bm k})]$.  As before, $\hat 1$, $\hat 2$, and $\hat 3$ are vectors of nearest-neighbor links on the triangular lattice. We find that the static spin susceptibility $\chi_{x}^0$ takes a nonzero value given by the integral over the Brillouin zone (BZ), Eq.~\eqref{eq:chilowT}. Its numerical value depends on the parameters $\Delta$, $\mu_x$, $\mu_z$, and on $s=\pm 1$. In the limit $\Delta\lesssim | \mu_x - \mu_z | \ll 1$, $\chi_{x}^0$ approaches the Pauli susceptibility of two unpaired Fermions, $\chi_x^0\simeq 2\nu_z$, where $\nu_z = \int_{\text{BZ}} d^2k/(2\pi)\, \delta(\xi_{\bm k}^z)$ is the density of states at the Fermi surface. 
In conclusion, we predict a strong anisotropy of the spin susceptibility $\chi_{a}^0 = \text{Re}[R_{aa}(0)]$ for the d+id state at low temperature. 

The nuclear spin relaxation rate is given by $T_1^{-1} \sim T \text{Im} [ R(i\omega\rightarrow 0) ]$. In the d+id QSL state, we find that this quantity is exponentially small for temperatures below the gap.


\subsection{Specific heat and Wilson ratio}

In the d+id spin liquid, the magnetic specific heat at low temperature is linear in temperature due to the $f_z$ spinon Fermi surface. The coefficient is given by\cite{mahan}
\begin{equation}
   \gamma = \frac{C_M}{T} = \frac{\pi^2\nu_z}{3}\, .
\end{equation}
The Wilson ratio is defined as
\begin{equation}\label{eq:wilson}
  R_W = \frac{4\pi^2}{3} \frac{\bar\chi_0}{\gamma}\,.
\end{equation}
Since the measurements in Ba$_3$NiSb$_2$O$_9$ were made on powder samples, a directional average should be used in this expression for comparison with experiment, $\bar\chi_0 = 2\chi_{x}^0/3$.

\begin{figure}
\includegraphics[width=.5\textwidth]{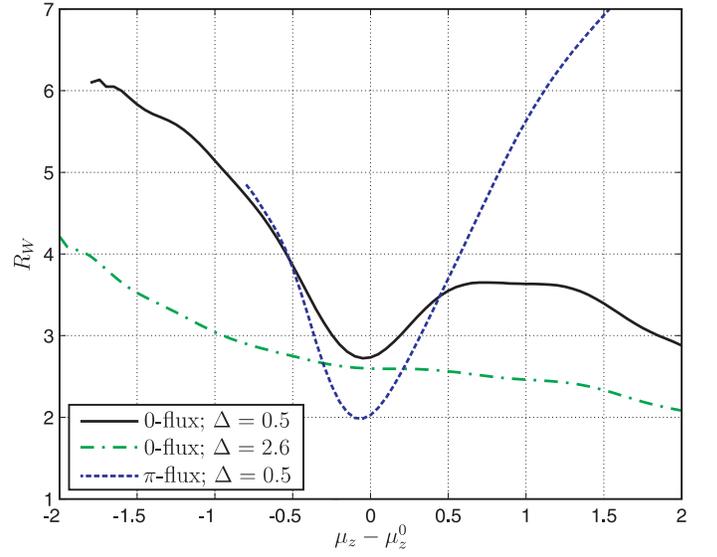}
\caption{Wilson ratio, \eqref{eq:wilson}, for the d+id state as a function of the spinon chemical potential, $\mu_z-\mu_z^0$. The shift $\mu_z^0$ corresponds to the optimal value of the chemical potential in the ring exchange model \eqref{eq:ringModel} at $\alpha=\pi/4$ (without single-ion anisotropy).}
\label{fig:wilson}
\end{figure}

The Wilson ratio, $R_W = 8 \chi_x^0 / (3 \nu_z)$, for the d+id state is plotted as a function of $\mu_z - \mu_z^0$ in Fig.~\ref{fig:wilson}. The choices of parameters ($\Delta=0.5$ and $2.6$ for the $0$-flux state, and $\Delta=0.5$ for the $\pi$-flux state) are examples of lowest energy d+id states in the ring-exchange model, \eqref{eq:ringModel}, at $\alpha\simeq\pi/4$. Note that, in this plot, we adjust the chemical potential $\mu_x = \mu_y$ such that the constraint is satisfied {\it on average}, $\sum_a \langle n_a \rangle = 1$. The shift $\mu_z^0$ is the optimized chemical potential for the ring-exchange model, i.e., for $D=0$. In Fig.~\ref{fig:wilson}, we see that the Wilson ratio is enhanced in the d+id state with respect to a metal (where $R_W=4/3$) by a factor of approximately two at $\mu_z = \mu_z^0$. This can be attributed to the fact that only a {\it single} fermion flavor contributes to the coefficient of specific heat in the QSL state. Since $S_{z}^2 = 1-f_{z}^\dagger f_z$, a single-ion anisotropy term in the Hamiltonian acts as a chemical potential for the $f_z$ spinon. We have $D \propto (\mu_z - \mu_z^0)$, and $R_W$ can be further enhanced by a non-zero $D$. An easy-plane anisotropy ($D<0$) shrinks the spinon Fermi surface, resulting in enhancement of $R_W$. For an easy-axis anisotropy ($D>0$), the Wilson ratio is enhanced due to an increase in magnetic susceptibility in the case of the $\pi$-flux state.

Experimentally, a large Wilson ratio of $R_W\simeq 5.6$ was reported for the spin liquid phase of Ba$_3$NiSb$_2$O$_9$.\cite{balicas11b} Within the framework of the d+id QSL state, we can conclude that quite a strong single-ion anisotropy, $|D|\gtrsim 1$, is required to explain the large Wilson ratio seen in Ba$_3$NiSb$_2$O$_9$.

Note that the U(1) state has Fermi surfaces for all three spinon flavors. However, since this state is in a Coulomb phase, the U(1) gauge fluctuations are expected to be very strong. Assuming Landau damping of the photon, it has been proposed that the specific heat in such a strongly coupled phase should have the non-Fermi-liquid behavior $C_M \propto T^{2/3}$.\cite{montrunichTr,leeLee05} 

\subsection{Thermal Hall effect}

Due to the spinon Fermi surface of $f_z$, the d+id QSL state exhibits a longitudinal heat conductivity.\cite{katsuraNagaosaLee09} According to the Wiedemann-Franz law, it is of the form
\begin{equation}
  \kappa^{xx} = \frac{ \tau\ \varepsilon_f}{\hbar}\, g_0\,,
\end{equation}
where $g_0 = \pi^2 T/(3 h)$ is the thermal conductance quantum, $\epsilon_f$ is the Fermi energy of the $f_z$ spinon, and $\tau$ is its lifetime. However, no longitudinal spin current will flow since the spin excitations are fully gapped in the bulk. Nevertheless, we expect a thermal (and spin) Hall conductivity due to the chiral edge modes:\cite{kaneFisher97,readGreen00}
\begin{equation}
  \kappa^{xy} \simeq 2 g_0\,.
\end{equation}
Since the state is compressible, $\kappa^{xy}$ is not expected to be exactly quantized. The $f_z$ spinon with a bulk Fermi surface also contributes to $\kappa^{xy}$ due to a classical Hall effect in the chiral spin liquid. On the other hand, the spin Hall conductivity is expected to be exactly quantized.


Let us briefly contrast the physical properties of the d+id QSL state discussed here with the spin liquid scenario proposed by Xu {\it et al.}\cite{xu12} for the 6H-B phase of the Ba$_3$NiSb$_2$O$_9$ compound. The proposed (``$Z_4$'') state has gapless fermionic spinon excitations with quadratic band touching. This leads to a $T$-linear specific heat and a constant spin susceptibility at low temperature. However, in contrast to the d+id QSL, the bulk spin excitations are gapless in the $Z_4$ state, and no chiral edge modes are expected. This leads to a finite spin relaxation rate at low temperature as well as absence of thermal and spin Hall effects in this state.

\section{Conclusion and outlook}\label{sec:conclude}

In this paper, we construct all natural quantum spin liquid states with three flavors of fermionic spinons for spin $S=1$ Heisenberg models on the triangular lattice. We compare their variational energies with the ones of various long-range ordered states. We find that for large biquadratic and ring-exchange terms (of the order of the Heisenberg exchange $J>0$), an exotic chiral quantum spin liquid with a spinon Fermi surface is stabilized. 
The physical properties of the d+id QSL state seem to be consistent with the recent experiment on Ba$_3$NiSb$_2$O$_9$.\cite{balicas11a}

While the d+id QSL scenario
we investigate in this paper has many attractive and novel features, it remains unclear whether the microscopic parameters required to stabilize such a phase are realized in Ba$_3$NiSb$_2$O$_9$. From the crystal structure proposed in [\onlinecite{balicas11b}], it seems more likely that the nearest-neighbor antiferromagnetic exchange energy $J$ is the dominant microscopic parameter. Therefore, the theoretical research must continue and more experiments are needed to elucidate the spin state realized in this material.

Recently, new experimental results were published on the related spin-liquid candidate Ba$_3$CuSb$_2$O$_9$ in [\onlinecite{nakatsuji12}]. In contrast to earlier experiments on powder samples,\cite{balicas11a} the new experiments on single crystals indicated that the Cu$^{2+}$ ions on the triangular lattice may form dipolar molecules with the Sb$^{5+}$ ions and can move out of plane. Strong disorder due to Jahn-Teller distortions or fluctuations of these Ising dipoles may play a key role in the absence of ordering in the Cu compound. Similar effects may also be present in Ba$_3$NiSb$_2$O$_9$, which opens promising avenues for future studies on this material.

\section*{Acknowledgements}

We thank Kuang-Ting Chen, Rebecca Flint, Dmitri Ivanov, Z.-X.~Liu, Tai-Kai Ng, Lara Thompson, Tam\'as T\'oth, and Fa Wang for helpful discussions. TS is supported by NSF DMR 1005434. PAL is supported by NSF DMR 1104498. SB acknowledges support from the Swiss National Science Foundation (SNSF).


\appendix
\section{Variational Monte Carlo}\label{app:VMC}

The Variational Monte Carlo (VMC) method allows to efficiently evaluate expectation values of observables in a given many body wave function within small error bars.\cite{ceperley77,gros88} This works as follows: Let $|\psi\rangle$ be the wave function and let $O$ be the observable we want to evaluate. Let $\{|\alpha\rangle\}$ be an ``Ising'' basis of the Hilbert space; i.e., $|\alpha\rangle$ is a product of local basis states. We can write
\begin{equation}
   \langle \psi | O |\psi\rangle = \sum_{\alpha} |\psi(\alpha)|^2 \frac{\langle\alpha|O|\psi\rangle}{\langle\alpha|\psi\rangle}\,,
\end{equation}
with $|\psi(\alpha)|^2 = |\langle\alpha|\psi\rangle|^2/\langle\psi|\psi\rangle$. Since $\sum_\alpha |\psi(\alpha)|^2 = 1$, $|\psi(\alpha)|^2$ is a probability distribution on the Ising configurations $\{\alpha\}$. Such a distribution can be generated by a Metropolis algorithm with acceptance probability
\begin{equation}\label{eq:metropolis}
  p(\alpha\rightarrow\alpha') = \text{min}\{\left|\frac{\psi(\alpha')}{\psi(\alpha)}\right|^2, 1 \} \, .
\end{equation}
Note that, in \eqref{eq:metropolis}, $\psi(\alpha)$ does not need to be normalized. The sequence $\{\alpha\}$ generated by a random walk with probability \eqref{eq:metropolis} can be used to efficiently calculate the expectation value,
\begin{equation}\label{eq:MCapprox}
  \langle \psi | O |\psi\rangle \simeq \sum_{\{\alpha\}} \frac{\langle\alpha| O |\psi\rangle}{\langle\alpha | \psi\rangle}\, .
\end{equation}
In this paper, we use $\sim 200$ Monte Carlo runs to estimate the error of Eq.~\eqref{eq:MCapprox} by its variance over the runs. The length of a run is $\sim 200$ steps, and the observables are measured after each step. The measurements are precessed by an equilibration skip of $\sim 400$ steps. Each Monte Carlo step consists of $4\times N\sim 600$ local moves, accepted with probability \eqref{eq:metropolis}. We use a lattice of $N=L\times L$ sites, with a linear size $L=12$ in our calculations. The error bars on the variational energies shown in Figs.~\ref{fig:KDmodel} and \ref{fig:Rex} are smaller than the symbol sizes.

Local and global constraints (projections) on the wave function $|\psi\rangle$ can be easily implemented in the VMC scheme. The Gutzwiller projection, $|\psi\rangle = P_G|\psi_0\rangle$, for example, can be taken into account by restricting the Ising configurations $\alpha$ to the singly occupied subspace ($n_j \equiv 1$). Similarly, projection of a spin wave function to $S_z^\text{tot}=0$ leads to a global restriction on the configurations $\alpha$.
Here, it is important to have an algorithm that generates all states $\alpha$ in the constrained subspace with uniform probability.

To apply VMC to a particular wave function, we first need an expression for $\psi(\alpha) \propto \langle\alpha|\psi\rangle$. Next, an efficient algorithm is needed to calculate the Metropolis acceptance probabilities \eqref{eq:metropolis} for local moves in the constrained subspace. Similarly, for each observable of interest, one has to find an efficient way to calculate the ratio of overlaps in \eqref{eq:MCapprox}.

\section{Fermionic wave functions}\label{app:fermionic}

The first class of wave functions that we are considering in this paper are Gutzwiller-projected ground states of quadratic Hamiltonians, $H_\text{MF}$, for three flavors of fermions $f_a$. A similar study of wave functions with {\it two} flavors of fermions has been pioneered by Gros\cite{gros88} for spin $S=1/2$ models.

In our calculation of fermionic QSL and fermionic ordered wave functions, we use the local basis of time-reversal invariant states, $\ket{a}\in\{\ket{x},\ket{y},\ket{z}\}$, Eq.~\eqref{eq:xbasis}. The Ising configurations $\alpha$ are restricted to singly occupied states on a lattice of $N = L\times L$ sites. Furthermore, we restrict the configurations to states with $N_x = N_y$ and $N_z$ kept fixed (that is, the wave functions are projected to fixed total flavor numbers; see below).

Let ${\bm r}_{j}^a\in \mathbb{Z}^L\times\mathbb{Z}^L$ be the lattice positions of flavor $a\in \{x, y, z\}$ in the Ising configuration $\alpha$. The U(1) state and the triplet ($x$-$y$ paired) QSL states in \eqref{eq:Hqsl} can be written as a product of two determinants,
\begin{equation}\label{eq:tripletwf}
  \langle\alpha|\psi\rangle = \text{det} [e^{i{\bm k}_{j}^z\cdot{\bm r}_{l}^z}]\, \text{det} [ A({\bm r}_{j}^x - {\bm r}_{l}^y)]\,,
\end{equation}
where $j$ and $l$ are the indices for the determinants. ${\bm k}_{j}^z$ are the occupied momentum states of $f_z$ spinons inside the Fermi sea, $\epsilon_{{\bm k}^z_j} < \mu_z$. For the U(1) state, $A({\bm r})$ is a Slater matrix,\cite{gros88}
\begin{equation}\label{eq:U1a}
  A({\bm r}) = \sum_{\substack{{\bm k}\in\text{BZ},\\ \epsilon_{\bm k} <\mu_x}} e^{i {\bm k}\cdot {\bm r}}\,,
\end{equation}
with momenta ${\bm k}$ going over filled states in the first Brillouin zone (BZ). For the triplet QSL states ($s$-wave, d+id), we have
\begin{equation}\label{eq:tripletwfa}
  A({\bm r}) = \sum_{{\bm k}\in\text{BZ}} a_{\bm k} \, e^{i {\bm k}\cdot {\bm r}}\,,
\end{equation}
where $a_{\bm k} = v_{\bm k}/u_{\bm k} = \Delta_{\bm k}/(E_{\bm k} + \xi_{\bm k})$ is the ratio of BCS coherence factors for the pairing of $f_x$ and $f_y$ fermions.\cite{anderson87}

For the QSL states with equal-flavor pairing ($f$-wave, p+ip), the wave function is a product of three Pfaffians,\cite{lhuillier88,bajdichWagner08}
\begin{equation}\label{eq:singletwf}
  \langle\alpha|\psi\rangle = \prod_a \text{Pf} [ A^a({\bm r}_{j}^a - {\bm r}_{l}^a)]\,,
\end{equation}
with
\begin{equation}\label{eq:singletwfa}
  A^a({\bm r}) = \sum_{{\bm k}\in \text{BZ}} a^a_{\bm k} \, \sin ({\bm k}\cdot {\bm r})\,,
\end{equation}
where $a^a_{\bm k} = v^a_{\bm k}/u^a_{\bm k}$ are the ratio of coherence factors for each paired fermion flavor.

In the case of the ordered states \eqref{eq:Hord}, 
the fermions are unpaired, but the flavors hybridize through terms $f_{ia}^\dagger f_{jb}$, etc. For a lattice of $N$ sites, the corresponding wave function is a single Slater determinant of size $N\times N$,
\begin{equation}\label{eq:orderedwf}
  \langle\alpha|\psi\rangle = \text{det} [ A_l({\bm r}_{j}^a) ]\,.
\end{equation}
Here, $l=1\ldots N$, and $A_l({\bm r}_j^a)$ are the lowest eigenvectors of the mean-field matrix $H_{ij}^{ab}$ with $H_\text{ord} = \sum_{i j, a b} f_{a i}^\dagger H_{ij}^{ab} f_{b j}$. [For the three-sublattice ordered states we consider in this paper, the eigenvectors can be labeled by $l=(n,{\bm k})$, where $n$ is a band index and $\bm k$ lies in the reduced Brillouin zone.]

Our calculations are done on a finite lattice with $N = L\times L$ sites. In order to avoid singularities or degeneracies in \eqref{eq:tripletwf}--\eqref{eq:orderedwf}, we use quadratic trial Hamiltonians \eqref{eq:Hqsl} and \eqref{eq:Hord} with periodic in one and antiperiodic boundary conditions in the other lattice direction for the spinons $f_{aj}$. The $f$-wave state, however, has lines of nodes in the gap function $\Delta_{\bm k}$ (at momenta $\{{\bm k}_0\}$) that cannot be avoided by choosing periodic-antiperiodic boundary conditions. A singularity $|a^a_{{\bm k}_0}| \rightarrow \infty$ occurs on these lines, and \eqref{eq:singletwfa} is ill defined. To cure the divergencies, we replace $a^a_{{\bm k}_0}$ by a large but finite quantity, namely, $\pm 20\times \text{max}_{{\bm k}\notin\{{\bm k}_0\}}|a^a_{\bm k}|$. The sign is chosen to be consistent with the sign of $a^a_{\bm k}$ as ${\bm k}\rightarrow {\bm k}_0$. We have verified that the relevant correlators do not depend on the precise factor in the regularization and that the wave function (correlators) correctly reproduces the U(1) state when $|\Delta^{aa}| \ll 1$.

We use the usual tricks for an efficient evaluation of the Metropolis acceptance probability \eqref{eq:metropolis} and the expectation values \eqref{eq:MCapprox} in fermionic wave functions: The inverse of the matrices in \eqref{eq:tripletwf}, \eqref{eq:singletwf}, and \eqref{eq:orderedwf} is stored and updated during the Monte Carlo random walk.\cite{ceperley77} This allows for efficient evaluation of determinants and Pfaffians with rows and/or columns replaced or removed.\cite{gros88,martinRandall99,bajdichWagner08} To update the inverse of an antisymmetric matrix with a row and column replaced, we use the Sherman-Morrison algorithm\cite{NumRec} twice, followed by antisymmetrization of the matrix. This procedure greatly improves the numerical stability of the update. The ``pfapack'' package by Wimmer\cite{wimmer11} is used for efficient evaluation of Pfaffians.

\subsection*{Flavor-number nonconservation \label{app:VMC:grand}}

An important technical difficulty with fermionic RVB wave functions for spin $S=1$ is that typical microscopic models such as \eqref{eq:KDmodel}, when written in terms of fermion operators, do not conserve the number of each fermion flavor separately. This issue is also present if we wish to represent the spin operator by more than three fermion flavors. We have $N_a = N - \sum_j S_{aj}^2$ and $[H_{KD}, N_a]\neq 0$, in general. Unlike in the case of spin-$1/2$, conservation of $S_z^{\text{tot}} = \sum_j S_{zj}$ does not imply conservations of flavor number. Note, however, that $N_a$ is conserved in the SU(3) model \eqref{eq:ringModel} or in the $KD$-model \eqref{eq:KDmodel} at $K=1$, where this issue does not arise. Writing \eqref{eq:KDmodel} with fermions, the terms not commuting with $N_a$ are
\begin{equation}\label{eq:dangerousKD}
  ( K - 1 ) \sum_{ab} f_{a i}^\dagger f_{b i} f_{a j}^\dagger f_{b j}\,,
\end{equation}
which vanish for $K=1$. In general, there is therefore no justification for using variational wave functions that are particle-number eigenstates. For such wave functions, the Ising configurations $\alpha$ in \eqref{eq:metropolis} must visit all possible total flavor numbers, with $\sum_a N_a = N$ kept fixed. In a brute force implementation, the determinants and Pfaffians in \eqref{eq:tripletwf} and \eqref{eq:singletwf} may need to change sizes during a Monte Carlo run, which implies a high computational overhead. Such a simulation has recently been done in the case of spin-one chains.\cite{liu12}

The problem is actually absent for the QSL states with a spinon Fermi surface. In this case, the wave function is an $N_z$ eigenstate. $N_x$ and $N_y$ do fluctuate in a paired state; nevertheless, $N_x = N_y$ and all expectation values of \eqref{eq:dangerousKD} vanish in this class of wave functions.
The difficulty is only present for equal-flavor paired QSL states ($f$-wave and p+ip) and for the ordered states \eqref{eq:Hord}. In these cases, the expectation value of \eqref{eq:dangerousKD} does not vanish (before or after Gutzwiller projection). The flavor numbers $N_a$ fluctuate independently of each other in these wave functions.

To resolve this issue, we can use the standard argument\cite{gros88} that relates grand-canonical and microcanonical RVB wave functions: The paired mean-field states are strongly peaked at some average flavor number $\tilde {\bm N}_0 = (\tilde N_x^0, \tilde N_y^0, \tilde N_z^0)$. This peak in flavor number may shift position to ${\bm N}_0$ after Gutzwiller projection, but it should still be present. Furthermore, the variance is expected to vanish in the thermodynamic limit, $\langle(N_a - N_a^0)^2\rangle/N^2 \sim 1/N$. Therefore, it is justified to work with microcanonical wave functions that are obtained by projecting the grand-canonical wave function, $|\psi\rangle$, to fixed total flavor numbers,
\begin{equation}
  |{\bm N}_0\rangle = P({\bm N}_0)|\psi\rangle\, .
\end{equation}

VMC calculation of expectation values of particle-number conserving operators within a microcanonical wave function is straightforward. However, off-diagonal operators such as \eqref{eq:dangerousKD} require some care.\cite{bieriIvanov07} As an example, let us consider the operator
\begin{equation}\label{eq:dangerousex}
  R_{xy} = f_{x i}^\dagger f_{y i} f_{x j}^\dagger f_{y j}\,.
\end{equation}
Its expectation value in the grand-canonical wave function can be approximated as
\begin{equation}\label{eq:offdiag}
  \langle \psi| R_{xy} |\psi\rangle \simeq \frac{\langle {\bm N}_0^+ | R_{xy} |{\bm N}_0\rangle}{\sqrt{\langle {\bm N}_0^+ | {\bm N}_0^+ \rangle\langle {\bm N}_0|{\bm N}_0 \rangle}}\,,
\end{equation}
with ${\bm N}_0^\pm = (N_x^0 \pm 2, N_y^0 \mp 2, N_z^0)$, and ${\bm N}_0$ is the average particle number in $|\psi\rangle$. 
In VMC, the right-hand side of Eq.~\eqref{eq:offdiag} cannot be calculated directly with the correct normalization. However, it is possible to calculate
\begin{equation}\label{eq:avg1}
  \frac{\langle {\bm N}_0^+ | R_{xy} |{\bm N}_0\rangle}{\langle {\bm N}_0|{\bm N}_0 \rangle} \qquad\text{and}\qquad
  \frac{\langle {\bm N}_0 | R_{xy} |{\bm N}_0^-\rangle}{\langle {\bm N}_0|{\bm N}_0 \rangle}
\end{equation}
within a single Monte Carlo run. Since the last average satisfies $\left|\langle{\bm N}_0 | R_{xy} |{\bm N}_0^-\rangle/\langle {\bm N}_0|{\bm N}_0 \rangle\right| \simeq \left|\langle {\bm N}_0^+ | R_{xy} |{\bm N}_0\rangle/\langle {\bm N}_0^+|{\bm N}_0^+ \rangle\right|$, the normalization factor can be calculated from the ratio of the two correlators in~\eqref{eq:avg1},
\begin{equation}\label{eq:gxy}
  g_{xy} = \frac{\langle{\bm N}_0|{\bm N}_0\rangle}{\langle{\bm N}_0^+|{\bm N}_0^+\rangle} \simeq \left|\frac{\langle {\bm N}_0 | R_{xy} |{\bm N}_0^-\rangle}{\langle {\bm N}_0^+ | R_{xy} |{\bm N}_0\rangle}\right|\, .
\end{equation}
Finally, the correctly normalized expectation value \eqref{eq:offdiag} is evaluated as
\begin{equation}
  \langle \psi| R_{xy} |\psi\rangle \simeq \sqrt{g_{xy}}\, \frac{\langle {\bm N}_0^+ | R_{xy} |{\bm N}_0\rangle}{\langle {\bm N}_0|{\bm N}_0 \rangle}\, .
\end{equation}

It is clear that for a given wave function, $g_{xy}$, \eqref{eq:gxy}, does not depend on the off-diagonal operator $R_{xy}$ (for example, $R_{xy}$ on different sites must give the same $g_{xy}$). This provides a nontrivial check of our code and we found that the renormalization factors $g_{ab}$ are indeed identical on different sites within error bars.

Of course, a particle-number projection $|{\bm N}\rangle$ is only a faithful representation of $|\psi\rangle$ if the flavor number ${\bm N}$ is sufficiently close to the average value ${\bm N}_0$ in the Gutzwiller projected wave function. Using ${\bm N}$ as a variational parameter (here with the restriction $N_x=N_y$) guarantees that the state $|{\bm N}_0\rangle \propto |\psi\rangle$ is among the variational wave functions. For the equal-flavor paired singlet wave functions, we found that the agreement between our optimal correlators and the ones calculated in the corresponding grand-canonical wave functions is very good.\cite{liuComm}

For spin $S=1/2$ systems, the investigation of (doped) RVB wave functions in the grand-canonical ensemble 
was pioneered by Yokohama and Shiba in Ref.~[\onlinecite{yokoyamaShiba88}]. These authors introduced a particle-hole transformation $c_{\downarrow}^\dagger\mapsto c_{\downarrow}$ that allows one to do fixed-particle VMC simulations. However, this trick does not easily generalize to spin-one. For spin-half RVB wave functions, the agreement between microcanonical and grand-canonical approaches was found to be very good. Note, however, that particle number renormalization by the Gutzwiller projector in grand-canonical wave functions leads to subtle effects that need to be taken into account if one wishes to apply the Gutzwiller approximation.\cite{edegGros05,andersonOng06,fuku08}

\section{Huse-Elser wave functions}\label{app:huseelser}

This appendix contains details regarding the implementation of trial wave functions of Huse-Elser type, generalized to the spin $S=1$ case. Similar to the case of spin $S=1/2$,\cite{huseElser88} our construction starts from an uncorrelated product-state wave function. Quantum correlations are introduced by applying Jastrow factors to the simple product state. The resulting wave function has two sets of variational parameters: parameters controlling the product-state, and Jastrow parameters responsible for the quantum correlations.

For the Huse-Elser wave functions, we use the local basis of $S_z$ eigenstates, i.e., the states $\ket{0}$, $\ket{1}$, and $\ket{\bar 1}$ with $S_z=0$, $1$, and $-1$, respectively. The corresponding basis of Ising configurations is denoted by  $\ket{\alpha} = \ket{1 1 0 \bar 1 0 \bar 1 1\ldots }$. As before, the singly occupied subspace corresponds to physical spin states. Furthermore, we project the wave functions to $S_z^\text{tot}=0$ by restricting to Ising states with $N_1 = N_{\bar 1}$. However, here we allow the total flavor numbers to fluctuate within this subspace.


In Ref.~[\onlinecite{thesisToth11}] the optimal three-sublattice product states for the bilinear-biquadratic model \eqref{eq:KDmodel} were calculated. It was found that the ordering patterns in this model are well captured by the antiferromagnetic and nematic states given in Eqs.~(\ref{eq:magord}) and~(\ref{eq:nemord}). In the basis of $S_z$ eigenstates, the wave function on $A$, $B$, and $C$ sublattices is given by
\begin{equation}\label{eq:abcWavef}\begin{split}
  \ket{A}
  &=
  \cos\eta\, \ket{0} + \kappa\frac{\sin\eta}{\sqrt{2}}\, (\ket{1}+\ket{\bar 1})\,,\\
  \ket{B},\ket{C}
  &=
  \cos\eta\, \ket{0} + \kappa\frac{\sin\eta}{\sqrt{2}}\, (e^{\mp\frac{2\pi i}{3}}\ket{1}+e^{\pm\frac{2\pi i}{3}}\ket{\bar 1})\,,
\end{split}\end{equation}
where $\eta$ is a variational parameter. $\kappa = 1$ corresponds to the antiferromagnetic, and $\kappa = i$ to the nematic product state. Using the Ising basis, the corresponding wave function
may be written as
\begin{equation}\label{eq:psi0}
  \ket{\psi_p} = \sum_{\alpha} e^{
  \tilde H_1}\ket{\alpha},
\end{equation}
where the sum goes over Ising states in the $S_z$ basis. The one-body operator $\tilde H_1$ accounts for different weights of $\ket{0}$, $\ket{1}$, and $\ket{\bar 1}$, as well as for site-dependent phase factors in the product-state wave function. For the particular case specified in Eq.~\eqref{eq:abcWavef}, it can be written in terms of $S_z$ operators as
\begin{equation} \label{eq:tildeH0}
  \tilde H_1 = \sum_j \{ \frac{2\pi i}{3} (\delta_{j\in C}-\delta_{j\in B}) S_{z j} + \log(\kappa \frac{\tan\eta}{\sqrt{2}}) S_{z j}^2\}\,.
\end{equation}
The Kronecker symbols $\delta_{j\in B}$ and $\delta_{j\in C}$ are nonzero only for sites $j$ belonging to the $B$ or $C$ sublattice, respectively.

The advantage of the rather complicated form \eqref{eq:psi0} for writing a simple product state is that quantum correlations can be built in easily by adding extra terms to $\tilde H_1$. We define
\begin{equation}\label{eq:psi}
  \ket{\psi} = \sum_{\alpha} e^{
  \tilde H}\ket{\alpha},
\end{equation}
where
\begin{equation} \label{eq:tildeH}
  \tilde H = \tilde H_1 + \tilde H_2 + \tilde H_3 + \ldots,
\end{equation}
and $\tilde H_2$, $\tilde H_3$, $\ldots$ denote many-body Jastrow factors. The correlated wave function \eqref{eq:psi} is easy to use in VMC, as long as $\tilde H$ is diagonal in the Ising basis $|\alpha\rangle$. In this paper, we only consider two-body correlation terms,
\begin{equation} \label{eq:tildeH2}
 \tilde H_2 = -\sum_{\langle i,j\rangle} \{ \beta (S_{z i} S_{z j}) + \gamma (S_{zi} S_{zj})^2 \}\, .
\end{equation}
In principle, in Eq.~(\ref{eq:tildeH2}), the sum can go over farther-neighbor lattice sites, and the variational parameters $\beta$ and $\gamma$ may depend on the distance between sites. However, inclusion of farther-neighbor correlations are expected to have a small effect on the ground state energy.\cite{huseElser88} Because of this, and also, in order to have a number of variational parameters that is similar to the number of parameters used for the spin liquid wave functions, we consider only nearest-neighbor Jastrow factors here.

The VMC algorithm can now be applied to Huse-Elser wave functions as outlined in Appendix~\ref{app:VMC}. The wave function is given by
\begin{equation} \label{eq:scalarAlphaPsi}
\langle\alpha|\psi\rangle = 
e^{\tilde H (\alpha)}\,,
\end{equation}
where $\tilde H (\alpha) = \langle\alpha|\tilde H|\alpha\rangle$. The Metropolis acceptance probability \eqref{eq:metropolis} and the expectation values \eqref{eq:MCapprox} are straightforward to calculate. In contrast to the case of fermionic wave functions, no determinants or Pfaffians need to be evaluated or updated for this.


In contrast to similar wave functions for spin $S=1/2$, an important subtlety arises here in the generation of the random walk. For $S=1/2$ and $S_z^\text{tot}=0$, the configurations $\alpha$ are restricted to states with an equal number of up and down spins. Therefore, the only admissible local Monte Carlo move is an exchange of two opposite spins. For $S=1$, due to presence of the nematic state $\ket{0}$ with $S_z=0$, more local moves are possible. The Hilbert space for $S=1$ and $S_z^\text{tot}=0$ can be written as a direct sum of orthogonal subspaces (``$N_1$-sectors'') with a fixed number $N_1=0\ldots N/2$ of sites in configuration $\ket{1}$. The dimension of each $N_1$-sector is $D(N_1)={N\choose{N_1}}{{N-N_1}\choose{N_1}}$. There exist two types of local moves in a random walk through the Ising configurations: those leaving $N_1$ intact and those changing $N_1$ and moving to a different $N_1$-sector. The algorithm generating the random walk has to be unbiased with respect to moves between different sectors such that each $N_1$-sector is visited with probability $p(N_1) = D(N_1)/\sum_{n=0}^{N/2}D(n)$. We have checked that such a distribution is accurately generated by the following procedure. We pick two sites at random and, depending on the states found on the sites, perform the following move:
\begin{enumerate}[(i)]
  \item $\ket{0}\ket{1}$ or $\ket{0}\ket{\bar 1}$: exchange the states.
  \item $\ket{1}\ket{\bar 1}$: exchange states or change to $\ket{0}\ket{0}$, each with probabilities $1/2$.
  \item $\ket{0}\ket{0}$: change state to $\ket{1}\ket{\bar 1}$.
  \item $\ket{1}\ket{1}$ or $\ket{\bar 1}\ket{\bar 1}$: pick two different sites that are occupied by unequal flavors and exchange them.
\end{enumerate}
In (iv), when the configurations $\ket{1}\ket{1}$ or $\ket{\bar 1}\ket{\bar 1}$ are encountered, it is important to find two flavors to exchange, thus not changing the $N_1$-sector. For example, if our algorithm rejected this case, and retried with another pair of sites, the random walk would be \emph{biased} with respect to the distribution $p(N_1)$, resulting in a higher probability for visiting sectors with smaller $N_1$.


\section{Symmetries of $KD$- and SU(3)-models}\label{app:sym}

In this appendix, we elaborate on the symmetry properties of the bilinear-biquadratic model \eqref{eq:KDmodel} and of the SU(3) ring-exchange model \eqref{eq:ringModel} investigated in this paper.


Let us first discuss the SU(3) symmetry of these models. Writing the Heisenberg exchange operator for spin $S=1$, Eq.~\eqref{eq:exchange}, in terms of the operators ${\bm f} = (f_x,f_y,f_z)$, we have
\begin{equation}\begin{split}
  \mathcal{P}_{ij} &= {\bm S}_i \cdot {\bm S}_j + ({\bm S}_i \cdot {\bm S}_j)^2 - 1\\
   &= \sum_{ab} f_{a i}^\dagger f_{b i} f_{b j}^\dagger f_{a j} = {\bm f}_i^\dagger\cdot ({\bm f}_i\cdot{\bm f}_j^\dagger) {\bm f}_j\, .
\end{split}\end{equation}
In this notation it is clear that $\mathcal{P}_{ij}$ is invariant under a global transformation ${\bm f}\mapsto A {\bm f}$ where $A$ is a general $3\times 3$ unitary matrix. However, as discussed previously, the transformation $f_a\mapsto e^{i\phi} f_a$ with the same phase for all flavors does not change the corresponding spin state. Therefore, the relevant spin symmetry is $SU(3) = U(3)/U(1)$, and we can take $A\in SU(3)$. Similar to the operators $f_a$ that create these states, the spin states $|a\rangle$ transform in the fundamental representation of the SU(3) symmetry, by matrix multiplication with $A$. To find the action of the symmetry on spin operators, let us define
\begin{equation}\label{eq:generators}
  \hat Q_\mu = \sum_{a b} f_a^\dagger \lambda^{ab}_{\mu} f_b\,,
\end{equation}
where $\lambda_{\mu} = (\lambda^{ab}_{\mu})$, $\mu=1\ldots 8$, are the Gell-Mann matrices, generators of SU(3). Using
\begin{equation}
  [\hat Q_\mu, f_a] = \sum_b\lambda^{ab}_\mu f_b\,,
\end{equation}
it is clear that
\begin{equation}\label{eq:ftransf}
  A {\bm f} = e^{i\, \text{ad}(\hat Q)} {\bm f} = e^{i\hat Q} {\bm f} e^{-i\hat Q}\,,
\end{equation}
for $A = \exp\{i\sum_\mu\alpha_\mu \lambda_\mu\}$ and $\hat Q = \sum_\mu \alpha_\mu \hat Q_\mu$. Therefore, the spin operators
\begin{equation}
  {\bm S} = -i {\bm f}^\dagger \wedge {\bm f}
\end{equation}
transform as
\begin{equation}
  {\bm S} \mapsto e^{i\hat Q} {\bm S} e^{-i\hat Q}
\end{equation}
under an SU(3) symmetry transformation.

Rather than explicitly writing down all eight generators of the SU(3) symmetry, Eq.~\eqref{eq:generators}, in spin language using the Gell-Mann basis, let us mention an equivalent set of generators. This set consists of the three spin rotation generators $S_a$ and the five independent quadrupolar operators $Q_{ab} = (S_a S_b + S_b S_a)/2 - 2/3\, \delta_{ab}$.\cite{pencLauchli}


The ring-exchange model, Eq.~\eqref{eq:ringModel}, is written in terms of Heisenberg exchange operators $\mathcal{P}_{ij}$. Therefore, it has the large SU(3) symmetry discussed above for all values of parameter $\alpha$. The $KD$-model, Eq.~\eqref{eq:KDmodel}, enjoys the SU(3) symmetry only at the special point $K=1$ and $D=0$ in parameter space (where it is equivalent to the ring-exchange model at $\alpha=0$). Moving away from this special point, for general $K$ but keeping $D=0$, the symmetry is reduced to SO(3) spin rotation symmetry, generated by ${\bm S}$. Finally, for $D\neq 0$, this symmetry is further reduced to U(1) spin rotation about the $z$ axis.

When we move away from the SU(3) symmetric point along the line $K=1$ and $D\neq 0$, the symmetry is reduced to SU(2) on that line. Clearly, the spin rotation symmetry is reduced to $S_z$ as $D\neq 0$. To find the remaining unbroken generators, we need to determine the SU(3) generators that commute with the biquadratic term $S_z^2$. These generators are $S_x S_y + S_y S_x$ and $S_x^2 - S_y^2$. Hence, \{$S_z$, $S_x S_y + S_y S_x$, $S_x^2 - S_y^2$ \} are the three generators of an SU(2) symmetry of the model \eqref{eq:KDmodel} on the line $K=1$.

Let us briefly discuss the symmetry reasons behind the degeneracy of the correlated AFM and the nematic states, \eqref{eq:magord} and \eqref{eq:nemord}, on the line $K=1$. In terms of spinon operators, the relevant symmetry generator is written as
\begin{equation}
  S_y^2 - S_x^2 = f_x^\dagger f_x - f_y^\dagger f_y\,.
\end{equation}
From \eqref{eq:ftransf}, we see that $x$ and $y$ states simply acquire an opposite phase under this transformation: $f_x\mapsto e^{i\varphi} f_x$, $f_y\mapsto e^{ -i\varphi} f_y$. It is easy to check that the magnetic state \eqref{eq:magord} is mapped to the spin-nematic state \eqref{eq:nemord} for $\varphi=\pi/2$, i.e., when $f_x \mapsto i f_x$ and $f_y \mapsto -i f_y$. Furthermore, it is clear that the hopping term in \eqref{eq:Hord} and the Jastrow factors in \eqref{eq:jastrow} are invariant under this transformation. Hence, the correlated ordered states are exactly mapped into each other by this transformation.



\begin{thebibliography}{99}

\bibitem{lee08}
P.~A.~Lee, Science {\bf 321}, 1306 (2008).

\bibitem{balents10}
L.~Balents, Nature (London) {\bf 464}, 199 (2010).

\bibitem{anderson72}
P.~W.~Anderson,
Mat.\ Res.\ Bull.\ {\bf 8}, 153 (1973).

\bibitem{anderson87}
P.~W.~Anderson,
Science {\bf 235}, 1196 (1987).

\bibitem{rmftReview04}
P.~W.~Anderson, P.~A.~Lee, M.~Randeria, M.~Rice, N.~Trivedi and F.~C.~Zhang,
J.~Phys.\ Cond.\ Mat.\ {\bf 16}, R755 (2004).

\bibitem{leeReview08}
P.~A.~Lee,
Rep.\ Prog.\ Phys.\ {\bf 71}, 012501 (2008);
P.~A.~Lee, N.~Nagaosa, and X.-G.~Wen,
Rev.\ Mod.\ Phys.\ {\bf 78}, 17 (2006).

\bibitem{kanoda03}
Y.~Shimizu, K.~Miyagawa, K.~Kanoda, M.~Maesato, and G.~Saito,
Phys.\ Rev.\ Lett.\ {\bf 91}, 107001 (2003).

\bibitem{yslee07}
J.~S.\ Helton, K.~Matan, M.~P. Shores, E.~A.\ Nytko, B.~M.\ Bartlett, Y.~Yoshida, Y.~Takano, A.~Suslov, Y.~Qiu, J.-H.~Chung, D.~G.~Nocera, and Y.~S.~Lee,
Phys.\ Rev.\ Lett.\ {\bf 98}, 107204 (2007).

\bibitem{maegawa08}
T.~Itou, A.~Oyamada, S.~Maegawa, M.~Tamura, and R.~Kato,
Phys.\ Rev.\ B {\bf 77}, 104413 (2008).

\bibitem{balicas11a}
H.~D.~Zhou, E.~S.~Choi, G.~Li, L.~Balicas, C.~R.~Wiebe, Y.~Qiu, J.~R.~D.~Copley, and J.~S.~Gardner,
Phys.\ Rev.\ Lett.\ {\bf 106}, 147204 (2011).

\bibitem{balicas11b}
J.~G.\ Cheng, G.\ Li, L.\ Balicas, J.~S.\ Zhou, J.~B.\ Goodenough, C.\ Xu, and H.~D.\ Zhou, Phys.\ Rev.\ Lett.\ {\bf 107}, 197204 (2011).

\bibitem{ashkroftMermin}
N.~W.~Ashcroft and N.~D.~Mermin, {\it Solid State Physics} (Rinehart and Winston, New York, 1976).

\bibitem{serbyn11}
M.~Serbyn, T.~Senthil, and P.~A.~Lee,
Phys.\ Rev.\ B {\bf 84}, 180403 (2011).

\bibitem{xu12}
C.~Xu, F.~Wang, Y.~Qi, L.~Balents, and M.~P.~A.~Fisher,
Phys.\ Rev.\ Lett.\ {\bf 108}, 087204 (2012).

\bibitem{liuNg10a}
Z.-X.~Liu, Y.~Zhou, and T.-K.~Ng,
Phys.\ Rev.\ B {\bf 81}, 224417 (2010).

\bibitem{liuNg10b}
Z.-X.~Liu, Y.~Zhou, and T.-K.~Ng,
Phys.\ Rev.\ B {\bf 82}, 144422 (2010).

\bibitem{chen12}
G.~Chen, M.~Hermele, and L.~Radzihovsky,
Phys.\ Rev.\ Lett.\ {\bf 109}, 016402 (2012).

\bibitem{noteS1}
Here we consider {\it fermionic} QSL wave functions for spin $S=1$. QSLs in terms of {\it bosonic} spin flip operators for $S>1/2$ have also been considered in the literature; see, e.g., Ref.~[\onlinecite{scharfenbergerThomaleGreiter11}], and references therein. A study of a doped spin-1/2 Heisenberg model on the triangular lattice in terms of fermionic RVB wave functions can be found in Ref.~[\onlinecite{weber06}].

\bibitem{pencLauchli}
K.~Penc and A.~M.~L\"auchli, in {\it Introduction to Frustrated Magnetism}, edited by C.\ Lacroix, F.\ Mila, and P.\ Mendels (Springer, 2011).

\bibitem{leeLee05}
S.-S.\ Lee and P.~A.~Lee,
Phys.\ Rev.\ Lett.\ {\bf 95}, 036403 (2005).

\bibitem{noteSym}
For example, an $x$-$y$ pairing with $\Delta_{ij}^{xy} = + \Delta_{ij}^{yx} \neq 0$, or a state with $\mu_x\neq\mu_z$ break spin-rotation symmetry around the $z$ axis.

\bibitem{gros88}
C.~Gros, Phys.\ Rev.\ B {\bf 38}, 931 (1988);
Ann.\ Phys.\ (NY) {\bf 189}, 53 (1989).

\bibitem{noteXY}
The case of $x\text{-}y$ pairing also allows for {\it on-site} $s$-wave pairing with a term $\Delta^{xy} f_{xj}f_{yj}$ in the trial Hamiltonian. However, we found that such a pairing term does not gain any variational energy in the models we consider.

\bibitem{Wen02}
X.-G.~Wen, Phys.\ Rev.\ B {\bf 65}, 165113 (2002).

\bibitem{noteMP}
In this work, we also considered QSL wave functions with mixed pairing symmetries, i.e., states that break lattice rotation symmetry. However, we find that such QSL states are always higher in energy than the rotation-invariant QSL states for the models we consider.

\bibitem{wenWilcekZee89}
X.~G.~Wen, F.~Wilczek, and A.~Zee,
Phys.\ Rev.\ B {\bf 39}, 11413 (1989).

\bibitem{fradkin79}
E.~Fradkin and S.~H.~Shenker,
Phys.\ Rev.\ D {\bf 19}, 3682 (1979).

\bibitem{huseElser88}
D.~A.~Huse and V.~Elser,
Phys.\ Rev.\ Lett.\ {\bf 60}, 2531 (1988).

\bibitem{tsunetsuguArikawa06}
H.~Tsunetsugu and M.~Arikawa,
J.\ Phys.\ Soc.\ Jpn.\ {\bf 75} 083701 (2006).

\bibitem{lauchli06}
A.~L\"auchli, F.~Mila, and K.~Penc,
Phys.\ Rev.\ Lett.\ {\bf 97}, 087205 (2006).

\bibitem{thesisToth11}
T.~T\'oth, EPFL PhD thesis No.~5037 (2011).

\bibitem{noteJ}
We always choose the spin Jastrow factors in the 120$^\circ$-AFM Huse-Elser wave function \eqref{eq:jastrow} to lie {\it perpendicular} to the ordering plane of the spins. That is, for $D<0$ (when the spins order in the $x$-$z$ plane) the Jastrow factor in Eq.~\eqref{eq:jastrow} should read $\exp\{-\beta S_{y i} S_{y j} - \gamma (S_{y i} S_{y j})^2\}$.

\bibitem{noteD}
We investigate the range $|D|<1.5$ in detail. But even greater values of $D$ do not seem to stabilize the QSL states.

\bibitem{noteConfinement}
The field-theory description of this phase consists of a U(1) gauge field interacting with gapped fermionic spinons. Integrating out the massive fermions leads to a pure compact U(1) gauge theory in 2+1 dimensions, which is expected to be confining at all couplings due to the instanton condensate.\cite{polyakovBook}

\bibitem{granathOslund03}
M.~Granath and S.~\"Ostlund, Phys.\ Rev.\ B {\bf 68}, 205107 (2003).

\bibitem{macdonaldGirvinYoshioka88}
A.~H.~MacDonald, S.~M.~Girvin, and D.~Yoshioka, Phys.\ Rev.\ B {\bf 37}, 9753 (1988).

\bibitem{zhangWang06}
G.-M.~Zhang and X.~Wang,
J.\ Phy.\ A: Math.\ Gen. {\bf 39}, 8515 (2006).

\bibitem{thouless65}
D.~J.~Thouless, Proc.\ Phys.\ Soc.\ {\bf 86}, 893 (1965).

\bibitem{misguich99}
G.~Misguich, C.~Lhuillier, B.~Bernu, and C.~Waldtmann,
Phys.\ Rev.\ B {\bf 60}, 1064 (1999);
W.~LiMing, G.~Misguich, P.~Sindzingre, and C.~Lhuillier,
Phys.\ Rev.\ B {\bf 62}, 6372 (2000).

\bibitem{montrunichTr}
O.~I.~Motrunich, Phys.\ Rev.\ B {\bf 72}, 045105 (2005); {\it ibid.} {\bf 73}, 155115 (2006).

\bibitem{lauchli05}
A.~L\"auchli, J.C.~Domenge, C.~Lhuillier, P.~Sindzingre, and M.~Troyer, Phys.\ Rev.\ Lett.\ {\bf 95}, 137206 (2005).

\bibitem{refSU2}
Note that the 120$^\circ$ nematic state only involves two out of three flavors. Therefore, in this phase, the SU(3) symmetry of the model is spontanously broken to SU(2). As a result, the three-site ring exchange term merely renormalizes the two-site Heisenberg term, and the ground state is the SU(2) N\'eel state on the triangular lattice. We thank A.~L\"auchli for this remark.

\bibitem{bauer12}
B.~Bauer, Ph.~Corboz, A.~M.~L\"auchli, L.~Messio, K.~Penc, M.~Troyer, and F.~Mila,
Phys.\ Rev.\ B {\bf 85}, 125116 (2012).

\bibitem{fazekasBook}
P.~Fazekas,
{\it Electron Correlation and Magnetism},
World Scientific (1999).

\bibitem{kittel60}
In C.~Kittel, Phys.\ Rev.\ {\bf 120}, 335 (1960), it was proposed that biquadratic terms can also arise from lattice distortions.

\bibitem{bastardis07}
R.~Bastardis, N.~Guih\'ery and C.~de Graaf,
Phys.\ Rev.\ B {\bf 76}, 132412 (2007).

\bibitem{hermele10}
A.~V.~Gorshkov, M.~Hermele, V.~Gurarie, C.~Xu, P.~S.~Julienne, J.~Ye, P.~Zoller, E.~Demler, M.~D.~Lukin, and A.~M.~Rey,
Nat.\ Phys.\ {\bf 6}, 289 (2010).


\bibitem{volovik97}
G.~E.~Volovik, Pis'ma Zh.\ \`Eksp.\ Teor.\ {\bf 66}, 492 1997 [JETP Lett.\
{\bf 66}, 522 (1997)].

\bibitem{senthilMarstonFisher99}
T.~Senthil, J.~B.~Marston, and M.~P.~A.~Fisher,
Phys.\ Rev.\ B {\bf 60}, 4245 (1999).

\bibitem{stoneRoy04}
M.~Stone and R.~Roy,
Phys.\ Rev.\ B {\bf 69}, 184511 (2004).

\bibitem{senguptaDasSarma01}
K.~Sengupta, I.~Zutic, H.-J.~Kwon, V.~M.~Yakovenko, and S.~Das Sarma,
Phys.\ Rev.\ B {\bf 63}, 144531 (2001).

\bibitem{ivanov04}
D.~A.~Ivanov,
Phys.\ Rev.\ Lett.\ {\bf 86}, 268 (2001).

\bibitem{takayamaLinMaki79}
H.~Takayama, Y.~R.~Lin-Liu, and K.~Maki,
Phys.\ Rev.\ B {\bf 21}, 2388 (1980).

\bibitem{wen91}
X.-G.~Wen, Phys.\ Rev.\ B {\bf 43}, 11025 (1991).


\bibitem{mahan}
G.~D.~Mahan, {\it Many Particla Physics} (Springer, 1990).

\bibitem{katsuraNagaosaLee09}
H.~Katsura, N.~Nagaosa, P.~A.~Lee,
Phys.\ Rev.\ Lett.\ {\bf 104}, 066403 (2010).

\bibitem{kaneFisher97}
C.~L.~Kane and M.~P.~A.~Fisher,
Phys.\ Rev.\ B {\bf 55}, 15832 (1997).

\bibitem{readGreen00}
N.~Read and D.~Green, Phys.\ Rev.\ B {\bf 61}, 10267 (2000).

\bibitem{nakatsuji12}
S.~Nakatsuji {\it et al.}, Science {\bf 336}, 559 (2012).

\bibitem{ceperley77}
D.~Ceperley and G.~V.~Chester,
Phys.\ Rev.\ B {\bf 16}, 3081 (1977).

\bibitem{lhuillier88}
J.~P.~Bouchaud, A.~Georges, and C.~Lhuillier,
J.\ Phys.\ France {\bf 49}, 553 (1988);
J.~P.~Bouchaud and C.~Lhuillier,
Europhys.\ Lett.\ {\bf 3}, 1273 (1987).

\bibitem{bajdichWagner08}
M.~Bajdich, L.~Mitas, L.~K.~Wagner, and K.~E.~Schmidt, Phys.\ Rev.\ B {\bf 77}, 115112 (2008);
M.~Bajdich, L.~Mitas, G.~Drobn\'y, L.~K.~Wagner, and K.~E.~Schmidt, Phys.\ Rev.\ Lett.\ {\bf 96}, 130201 (2006).

\bibitem{martinRandall99}
R.~A.~Martin and D.~Randall, in RANDOM-APPROX~'99, p.\ 257$-$268, Lecture Notes in Computer Science (Springer, 1999).

\bibitem{NumRec}
W.~H.~Press, S.~A.~Teukolsky, W.~T.~Vetterling, and B.~P.~Flannery, {\it Numerical Recipes in C} (Cambridge University Press, 1992).

\bibitem{wimmer11}
M.~Wimmer, http://arxiv.org/abs/1102.3440 (2011); ACM Trans.\ Math.\ Software {\bf 38}, 30 (2012).

\bibitem{liu12}
Z.-X.~Liu, Y.~Zhou, H-H.~Tu, X.-G.~Wen, and T.-K.~Ng, Phys.\ Rev.\ B {\bf 85}, 195144 (2012).

\bibitem{bieriIvanov07}
S.~Bieri and D.~Ivanov,
Phys.\ Rev.\ B {\bf 75}, 035104 (2007).

\bibitem{liuComm}
Z.-X.~Liu, private communications.

\bibitem{yokoyamaShiba88}
H.\ Yokoyama and H.\ Shiba,
J.~Phys.\ Soc.\ Jpn.\ {\bf 57}, 2482 (1988).

\bibitem{edegGros05}
B.~Edegger, N.~Fukushima, C.~Gros, and V.~N.~Muthukumar,
Phys.\ Rev.\ B {\bf 72}, 134504 (2005).

\bibitem{andersonOng06}
P.~W.~Anderson and N.~P.~Ong,
J.~Phys.\ Chem.\ Solids {\bf 67}, 1 (2006).

\bibitem{fuku08}
N.~Fukushima,
Phys.\ Rev.\ B {\bf 78}, 115105 (2008).


\bibitem{scharfenbergerThomaleGreiter11}
B.~Scharfenberger, R.~Thomale, and M.~Greiter,
Phys.\ Rev.\ B {\bf 84}, 140404 (2011).

\bibitem{weber06}
C.~Weber, A.~Laeuchli, F.~Mila, and T.~Giamarchi,
Phys.\ Rev.\ B {\bf 73}, 014519 (2006).

\bibitem{polyakovBook}
A.~M.~Polyakov, Nucl.\ Phys.\ {\bf B120}, 429 (1977); {\it Gauge Fields and Strings} (Harwood Academic Publishers, 1987).

\end{thebibliography}
\end{document}